\documentclass[aps,prb,twocolumn,eqsecnum,floatfix,showpacs,superscriptaddress]{revtex4}
\usepackage[dvips]{graphicx}
\usepackage[usenames,dvipsnames]{xcolor}
\usepackage{amsmath,amssymb,amsbsy,bm}
\usepackage{color,epsf,epsfig}

\newcommand{\ket}[1]{|{#1} \rangle}
\newcommand{\bra}[1]{\langle {#1}|}
\newcommand{\bs}{{\mathbf s}}
\newcommand{\mean}[1]{\langle{#1}\rangle}

\begin{document}
\preprint{}
\title[]{Effect of Randomness on Quantum Data Buses of Heisenberg Spin Chains}
\author{Sangchul Oh}
\affiliation{
Department of Physics, University at Buffalo, State University of New York,
Buffalo, New York 14260-1500, USA}
\author{Yun-Pil Shim}
\affiliation{Department of Physics, University of Wisconsin-Madison,
Madison, Wisconsin 53706, USA}
\author{Jianjia Fei}
\affiliation{Department of Physics, University of Wisconsin-Madison,
Madison, Wisconsin 53706, USA}
\author{Mark Friesen}
\affiliation{Department of Physics, University of Wisconsin-Madison,
Madison, Wisconsin 53706, USA}
\author{Xuedong Hu}
\thanks{Corresponding Author; email:xhu@buffalo.edu}
\affiliation{
Department of Physics, University at Buffalo, State University of New York,
Buffalo, New York 14260-1500, USA}
\date{\today}

\begin{abstract}
A strongly coupled spin chain can mediate long-distance effective couplings or entanglement 
between remote qubits, and can be used as a quantum data bus. We study how the fidelity of 
a spin-1/2 Heisenberg chain as a spin bus is affected by static random exchange couplings 
and magnetic fields. We find that, while non-uniform exchange couplings preserve the isotropy 
of the qubit effective couplings, they cause the energy levels, the eigenstates, and 
the magnitude of the couplings to vary locally. On the other hand, random local magnetic 
fields lead to an avoided level crossing for the bus ground state manifold, and cause 
the effective qubit couplings to be anisotropic. Interestingly, the total magnetic moment 
of the ground state of an odd-size bus may not be parallel to the average magnetic field. 
Its alignment depends on both the direction of the average field and the field distribution, 
in contrast with the ground state of a single spin which always aligns with the applied 
magnetic field to minimize the Zeeman energy. Lastly, we calculate sensitivities of the spin 
bus to such local variations, which are potentially useful for evaluating decoherence when 
dynamical fluctuations in the exchange coupling or magnetic field are considered.
\end{abstract}
\maketitle

\section{Introduction}
\label{sec:introduction}

A qubit is the elementary unit of quantum information, and can be realized with a variety of 
two-level systems, such as confined electron spins in a semiconductor nanostructure. For electron 
spin based qubits, universal quantum gates can be realized using Zeeman coupling and spin-spin 
exchange interaction.~\cite{LossPRA98} The direct Heisenberg exchange coupling between two electron 
spins is determined by the overlap of electron orbitals, and is thus a short-range nearest neighbor 
interaction. In order to implement quantum algorithms efficiently, quantum gate operations on 
remote qubits, i.e., controllable long-range couplings, are needed. Various quantum data buses 
have been introduced to bridge this gap.~\cite{Cirac95,Blais04,Bose03} In this context, we have 
proposed to use the ground states of a strongly coupled spin chain as a quantum data bus, or 
a spin bus.~\cite{Friesen07}  We have shown that the parity of the spin bus can significantly alter 
the long-range effective couplings and entanglement between qubits that are coupled to the spin 
bus,~\cite{Oh10} and external fields can modify the form of the effective interaction between 
the attached qubits.~\cite{Shim11} More recently, we have also shown that high-fidelity quantum 
state transfer can be achieved via such a spin bus.~\cite{Oh11}

An ideal quantum information processor has identical qubits, with precise control over couplings 
between qubits, and the qubits should be well isolated from their environment. However, in reality 
it is essentially impossible to create identical qubits based on artificial structures such 
as quantum dots and Josephson junctions, and in a solid state environment there are normally 
several sources of qubit variance. For example, the size of a quantum dot and the electron 
orbitals are largely determined by the gate structure and the applied gate voltages. They
can also be strongly influenced by factors such as the band structure of the host semiconductor 
and the random potential landscape due to modulation doping. Furthermore, the Coulomb exchange 
coupling between spin qubits is determined by the exponentially small overlap of the electron 
orbitals, and controlled by the gate voltages. Small variations in gate voltages could thus 
cause large changes in the exchange coupling. Such deviations from the ideal value could lead 
to imperfect gate operations, and possible gate errors.~\cite{Oh02} There are generally also 
very slow charge traps in a semiconductor heterostructure, where a trap can switch between 
two different charge distributions at a time scale much longer than the qubit operation time 
scales. While such a trap would probably be static during a quantum operation, it could modify 
the exchange coupling to a value that is different from the calibrated value. Similarly, via 
hyperfine interaction, environmental nuclear spins produce a local random magnetic field for 
a quantum dot confined electron spin qubit.~\cite{Merkulov02} This field can be considered 
quasi-static in the context of a spin bus because its dynamics is much slower than the bus 
mediated gates. In short, in building a practical quantum information processor, deviations 
from calibrated values for various control parameters are inevitable. It is thus necessary 
to know the engineering tolerance in the variation of parameters such as the spin-spin 
coupling and external magnetic fields.

In this paper we study how the capabilities of a spin bus are affected by static random 
variations in the exchange couplings between the bus node spins and the external magnetic 
fields experienced by the bus nodes. Specifically, we study how the bus spectrum, bus-qubit 
coupling, and bus-mediated qubit-qubit coupling are affected by these random but static 
variations of the system parameters.
The paper is organized as follows. In Sec.~\ref{sec:review} we discuss how the strongly 
coupled Heisenberg chain can act as a spin bus when external qubits are weakly attached 
to it. We derive the effective Hamiltonians of the qubit-bus system up to second order. 
In Sec.~\ref{sec:random}, we show how the fluctuations in exchange couplings and external 
magnetic fields within the spin bus could affect the fidelity of the bus. Finally, the summary 
and discussion are given in Sec.~\ref{sec:summary}. In the Appendices we discuss how to obtain 
the effective Hamiltonians using a projection method, and present more detailed results on 
the bus spectrum.

\section{Spin Chains as Quantum Data Buses: The Qubit-Bus Effective Hamiltonians}
\label{sec:review}

In this section, we discuss how a strongly coupled uniform antiferromagnetic Heisenberg 
spin chain can be used as a quantum data bus, or spin bus, which coherently connects remote 
qubits.~\cite{Friesen07,Oh10,Shim11,Oh11} In Appendix~\ref{app:eff_hamil}, using a many-body 
perturbation method based on the projection operator, we derive the effective Hamiltonians 
for the qubit-qubit and qubit-bus couplings to first and second order. We also calculate 
numerically relevant energy gaps, local magnetic moments, and the effective couplings for 
finite spin chains.

\begin{figure}[htbp]
\vspace{10pt}
\includegraphics[width=0.45\textwidth]{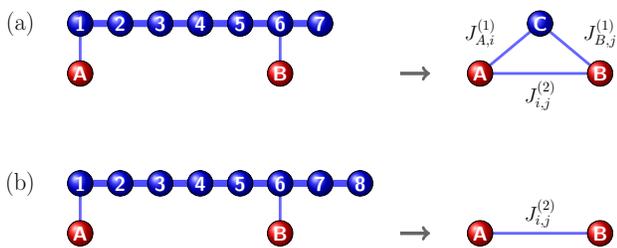}
\caption{(Color online) Left panel: Two external qubits $A$ and $B$ (red circles) are weakly 
coupled to an (a) odd-size or (b) even-size spin bus (blue circles). The right-hand panels 
show the corresponding effective Hamiltonians in the low energy limit. An odd-size spin bus 
in its ground state acts as an effective spin-$1/2$ particle denoted by $C$. It is coupled to 
the external qubits with strength $J^{(1)}_{\alpha i}$ at first order in the perturbation theory, 
and induces an RKKY-like coupling $J^{(2)}_{i,j}$ between the external qubits at second order. 
The even-size spin bus only mediates an RKKY-like coupling between the external qubits at 
second order.}
\label{Fig1}
\end{figure}

The system we consider has two spin qubits $A$ and $B$ weakly attached to 
an open spin chain $C$, as illustrated in Fig.~\ref{Fig1}. 
The Hamiltonian of the total system~\cite{Friesen07,Oh10,Shim11} is
\begin{align}
H = H_C + H_{QC} + H_Q \,.
\label{total_Hamil}
\end{align}
In the ideal case with uniform exchange couplings and a uniform external magnetic field 
(along the $z$ direction, ${\bf B}_0 = B_0\,{\bf\hat{z}}$), the Heisenberg Hamiltonian $H_C$ 
of the spin-$1/2$ chain is written as
\begin{align}
H_C = J_0\sum_{i=1}^{N-1}\bs_i\cdot\bs_{i+1}
    - g\mu_B B_0\sum_{i=1}^{N} s_{iz}\,,
\label{chain_Hamil}
\end{align}
where $\bs_{i}$ is the spin operator of the $i$-th node of the chain, $N$ is the total 
number of spins in the chain, and $J_0 > 0$ indicates a uniform antiferromagnetic 
coupling between any two nearest neighbor spins at sites $i$ and $i+1$. In this work, 
we will assume that a small or vanishing external magnetic field $B_0$ is applied to the spin 
chain (The operation of a spin chain under a finite external magnetic field 
is discussed in Refs.~\onlinecite{Shim11} and \onlinecite{Shim12}.).
Hereafter, the spin chain $C$ will be referred to as a spin bus because it acts as 
a quantum data bus.

The Hamiltonian $H_{QC}$ describes antiferromagnetic couplings of qubits $A$ and $B$ 
to the $i$-th spin and the $j$-th spin of the chain, respectively
\begin{align}
H_{QC} = J_{A,i}\,\bs_i \cdot{\bf S}_A + J_{B,j}\,\bs_j \cdot{\bf S}_A \,.
\end{align}
We assume that the bus-qubit couplings $J_{\alpha,i}$ with 
$\alpha=A,B$ are small enough that the spin bus remains in its ground state manifold 
at all times. In this limit, the total Hamiltonian $H$ can be split into 
the unperturbed Hamiltonian $H_0 = H_C + H_Q$ and the perturbation $H_1=H_{QC}$. 
The perturbation condition is 
\begin{equation}
J_{\alpha,i}/\Delta \sim NJ_{\alpha,i}/\pi^2 J_0 \ll 1 \,,
\label{perturbation_condition}
\end{equation}
where $\Delta \sim \pi^2 J_0/N$ is the zero-field gap~\cite{Lieb61} above the ground state 
manifold (for an odd-size bus) or ground state (for an even-size bus). 
$J_{\alpha,i}/\Delta$ is thus used as a perturbation parameter.  The qubit-bus coupling 
$J_{\alpha,i}$ can be turned on and off (gradually), unlike the static intra-bus coupling 
$J_0$. In general, external magnetic fields ${\bf B}_A$ and ${\bf B}_B$ may be applied to
qubits $A$ and $B$ to implement single-qubit operations on them, so that the Hamiltonian 
$H_{Q}$ can be written as
\begin{align}
H_{Q} = -g\mu_B{\bf B}_A\cdot{\bf S}_A -g\mu_B{\bf B}_B\cdot{\bf S}_B\,.
\end{align}
There is no direct exchange coupling between qubits $A$ and $B$, as they are nominally 
well separated. As shown later, the spin bus $C$ can mediate an effective coupling between 
them if they are both coupled to the bus. Since single-qubit operations are generally done 
separately from two- or multi-qubit operations, we set $H_Q = 0$ throughout this paper
and will focus on qubit-bus and two-qubit couplings.
Note that we will set $\hbar =1$ and $J_0 =1$ below for convenience.

In the perturbative limit described by Eq.~(\ref{perturbation_condition}), 
we can perform a canonical transformation of the full Hamiltonian in Eq.~(\ref{total_Hamil}) 
to obtain an effective Hamiltonian where the spin bus is in its ground state manifold.  
Details of the transformation are provided in Appendix A, as well as Refs.~\onlinecite{Shim11} 
and \onlinecite{Oh11}. The actual form of the effective interaction depends on the parity 
of the bus. An odd-size spin bus has a doubly degenerate ground manifold, and acts as 
an effective spin-1/2 particle. At first order in the perturbation theory, the spin-1/2 bus 
couples directly to the external qubits. At second order, the spin bus mediates 
an effective RKKY-like coupling between the qubits. The resulting effective Hamiltonian 
is given as follows:~\cite{Friesen07,Oh10,Oh11,Shim11}
\begin{align}
H_{\rm eff}^{(2)} 
= J^{(1)}_{A,i}\,{\bf S}_A\cdot{\bf S}_C 
+ J^{(1)}_{B,j}\,{\bf S}_B\cdot{\bf S}_C 
+ J^{(2)}_{i,j}\,{\bf S}_A\cdot{\bf S}_B \,,
\label{eff_Hamil_odd}
\end{align}
where ${\bf S}_C$ is a spin operator representing the ground doublet states of
the spin bus. The effective coupling between qubit $\alpha$ and the spin-bus $C$
is given to first order in the perturbation parameter by
\begin{subequations}
\label{Eq:coupling_J1}
\begin{align}
J_{\alpha,i}^{(1)} &\equiv J_{\alpha,i}\, m_i  \,,\\
m_i &=  \bra{0; \tfrac{1}{2}} \sigma_{iz} \ket{0;\tfrac{1}{2}}
     = -\bra{0;-\tfrac{1}{2}} \sigma_{iz} \ket{0;-\tfrac{1}{2}} \nonumber\\
    &=  \bra{0; \tfrac{1}{2}} \sigma_{ix} \ket{0;-\tfrac{1}{2}} \,.
\end{align}
\end{subequations}
This is a product of the bare coupling $J_{\alpha,i}$ between the $i$-th 
spin of the spin bus and the external qubit $\alpha$ and 
the expectation value $m_i$ of $\sigma_{iz}$ at site $i$ in the ground state 
of the spin bus. Notice that although $m_i$ is dimensionless, 
it can be considered as the local magnetic moment at site $i$ when multiplied 
by $g\mu_B/2$.
The RKKY-like second-order coupling $J^{(2)}$ is given by~\cite{RKKY}
\begin{align}
J^{(2)}_{i,j} \equiv \frac{J_{A,i} J_{B,j}}{2} \sideset{}{'}\sum_{n} 
			\frac{\bra{0}\sigma_{i\mu}\ket{n}\bra{n} \sigma_{j\mu}\ket{0}}{E_0 -E_n}\,.
\label{Eq:J_RKKY}
\end{align}
Here $E_n$ and $\ket{n}$ are the eigenenergies and eigenstates of $H_C$ of 
an isolated spin bus, and $\sigma_{i\mu}$ with $\mu=x,y,z$ stand for Pauli
operators of the $i$-th spin of the spin chain. The prime symbol on the summation 
indicates the exclusion of the ground states. 
At zero external field,
the ground state $\ket{0}$ in Eq.~(\ref{Eq:J_RKKY}) can be either $\ket{0;\tfrac{1}{2}}$ 
or $\ket{0;-\tfrac{1}{2}}$, or any linear combination between them. 
The choice does not change the value of $J^{(2)}_{i,j}$.
At a finite magnetic field, however, if the ground state of the spin bus is degenerate, the two
states are generally not spin-flipped image of each other.
In this case, Eq.~(\ref{Eq:J_RKKY}) has to be modified.

The effective Hamiltonian~(\ref{eff_Hamil_odd}) shows that the odd-size bus at zero or low 
field acts as an effective spin-$1/2$ particle that is coupled to the external qubits 
$A$ and $B$, as illustrated in Fig.~\ref{Fig1}.  Although in general 
$J^{(1)}_{\alpha,i} \gg J^{(2)}_{i,j}$, the second order term plays an essential role 
in long-time evolutions, such as in quantum state transfer.~\cite{Oh11} 
Thus our calculations in the rest of this paper are mostly concerned 
with these two coupling strengths.  Furthermore, we focus on their normalized form
$m_i = J^{(1)}_{\alpha,i}/J_{\alpha,i}$ and $K_{i,j}\equiv J^{(2)}_{i,j} J_0/J_{A,i}J_{B,j}$, 
which depend only on the size $N$ of the spin bus and the external magnetic field.

For an even-size bus, the sub-Hilbert space of interest is spanned by the non-degenerate 
ground state of the bus and the four eigenstates of the two qubits (again we focus on 
the low-field limit). Within this space the bus does not have any dynamics as it is 
represented by a single ground state. As for the two qubits, there is no first order 
effective coupling between them here, in contrast to the case of an odd-size bus. 
The second-order qubit coupling term is obtained in the same way as for an odd-size bus. 
The effective Hamiltonian to second order in the perturbation is given by~\cite{Oh10}
\begin{align}
H_{\rm eff}^{(2)} = J^{(2)}_{i,j}\,{\bf S}_A\cdot{\bf S}_B\,,
\end{align}
where the RKKY-like coupling $J^{(2)}_{i,j}$ has the same form as Eq.~(\ref{Eq:J_RKKY}). 
In this case, the prime indicates that the non-degenerate ground state is excluded from 
the summation.

Based on the effective Hamiltonians and the corresponding parameters, we can make some 
qualitative observations on where the bus-qubit system might be susceptible to randomness 
and fluctuations. In the case of an odd-size spin bus with attached qubits, 
the key features that determine the operation of the bus include the Zeeman splitting of 
the ground doublet of the spin bus, and the energy gap separating the ground doublet and 
the excited states, $\Delta_{12}$. 
The former depends on the magnetic environment for the bus, while the latter 
depends on the interaction strength between the bus nodes. Both the qubit-bus couplings 
$J^{(1)}_{\alpha,i}$ and the effective qubit-qubit couplings $J^{(2)}_{i,j}$
depend on the local exchange couplings and the local magnetic moments of the bus in 
its ground state manifold, which is a function of both magnetic environment and 
the intra-bus exchange couplings.  In the case of an even-bus with attached qubits, 
$J^{(2)}_{i,j}$ has similar dependence on system environment as in the 
odd-size bus case, and is thus susceptible to variations in both the local magnetic fields 
and exchange couplings.

\section{Effects of Randomness}
\label{sec:random}

In Sec.~\ref{sec:review}, we have shown how a Heisenberg spin chain with uniform exchange 
coupling $J_0$ acts as a spin bus. Now we address the main question of the present paper, 
on how static randomness in exchange couplings and external magnetic fields can affect 
the fitness of the spin chain as a quantum data bus. More specifically, we investigate how such 
randomness influences the two energy gaps, $\Delta_{01}$ and $\Delta_{12}$, and the effective 
qubit-bus and qubit-qubit couplings $J^{(1)}_{\alpha,i}$ and $J^{(2)}_{i,j}$.

In order to take into account the effects of randomness in exchange couplings and applied magnetic 
fields, the Hamiltonian of the chain (\ref{chain_Hamil}) is generalized to
\begin{align}
H_C = \sum_{i=1}^{N-1} J_i\,\bs_i\cdot\bs_{i+1} - g\mu_B\sum_{i=1}^{N} B_i\,s_{iz}\,,
\label{Hamil_random}
\end{align}
where $J_i > 0$ is the antiferromagnetic coupling between two neighboring spins at sites $i$ 
and $i+1$, and $B_i$ is the local magnetic field at the $i$th site of the spin bus. 
Note that in spite of the random $J_i$ and $B_i$, it can be easily shown that Hamiltonian 
(~\ref{Hamil_random}) still commutes with the $z$ component of the total spin 
${\bf S}$, i.e., $[H_C,S_z]=0$.

Hamiltonian~(\ref{Hamil_random}) may be considered as a finite quantum spin glass 
model.~\cite{Das} There are several spin glass models depending on the types
of couplings (Ising or Heisenberg, and short range or long range) and the distributions 
of $J_i$. For example, the Sherrington-Kirkpatrick model~\cite{SK75} has 
the couplings between arbitrary pairs, sampled from the normal distribution with zero mean,
while the Edwards-Anderson model~\cite{EA75} has only the nearest neighbor couplings.
In the context of quantum information processing, we can reasonably assume 
that the exchange couplings and the applied magnetic fields are both near their target values, 
$J_0$ and $B_0$. The random exchange couplings $J_i$ and applied magnetic fields $B_i$ are 
then
\begin{subequations}
\label{random_J&B}
\begin{align}
J_i &= J_0 + \delta J_i \,,\\
B_i &= B_0 + \delta B_i \,.
\label{random_B}
\end{align}
\end{subequations}
In the numerical analysis described below, we choose $\delta J_i$ and $\delta B_i$ 
that are randomly sampled from normal distributions with standard deviations $\sigma_J$ 
and $\sigma_B$, respectively. In experimental systems, we would expect such variations 
to be small, assuming reasonable calibration efforts. In the following studies, we analyze 
these two types of random variations separately, keeping one of the variables uniform.

\subsection{Effects of Random Exchange Couplings in Odd-size Buses}

In this subsection, we investigate how random variations in the inter-node exchange couplings 
$J_i$, given by Eq.~(\ref{random_J&B}a), affect the ability of an odd-size chain to function 
as a spin bus. Such variations could result from calibration errors, slow but random hopping 
of charge traps near the spin bus nodes, and whatever other factors that are not accounted for 
during the calibration process. Here the external magnetic field $B_0$ on the chain is set to 
be zero or small, so that the system remains in the isotropic regime. The external magnetic 
field, if any, is taken to be uniform, so that $\delta B_i=0$.

The variations $\delta J_i$ are small compared to $J_0$, so that they may be treated as 
a perturbation:
\begin{subequations}
\label{Eq:Hamil_random_J}
\begin{align}
H_C &= H_C^{(0)} + V \,,
\end{align}
where $H_C^{(0)}$ is the unperturbed Hamiltonian~(\ref{chain_Hamil}), and the perturbation
$V$ is given by
\begin{align}
V = \sum_{i=1}^{N-1} \delta J_i\,\bs_i\cdot\bs_{i+1}\,.
\end{align}
\end{subequations}
Hereafter the superscript $^{(0)}$ is used to denote the case of uniform exchange coupling 
or uniform magnetic fields in the bus. 
 
\begin{figure}[htb]
\includegraphics[scale=1.00]{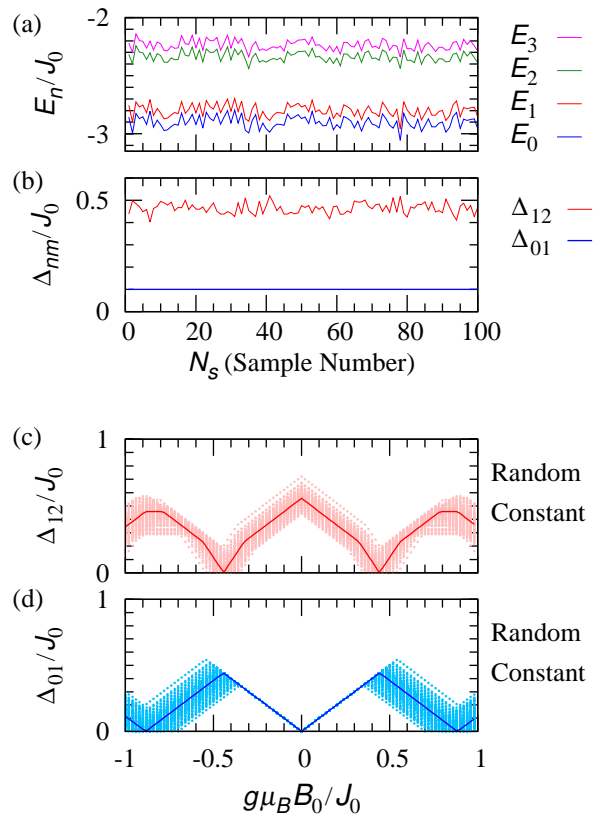}
\caption{(Color online) (a) Four lowest energy levels, $E_0, E_1,E_2$ and $E_3$, and (b) 
two energy gaps, $\Delta_{01} = E_1-E_0$ and $\Delta_{12}=E_2-E_1$ are plotted 
as a function of sample number $N_S$ of odd-size chains with size $N=7$ with random exchange 
couplings $J_i$ with the standard deviation $\sigma_J/J_0=0.1$ at $g\mu_BB_0/J_0 =0.1$.
In (c) and (d), two energy gaps $\Delta_{01}$ and 
$\Delta_{12}$ are plotted as a function of $B_0$ for the samples $N_s =100$.
}
\label{Fig:Odd_RJ_Gap}
\end{figure}

Here we consider an ensemble of $N_S$ odd-size chains.  Each of the samples in this ensemble 
has the same size $N$ but different $J_i$ sampled from the normal distribution with average 
$J_0$ and standard deviation $\sigma_J$. Fig.~\ref{Fig:Odd_RJ_Gap}(a) shows the fluctuations 
in energy levels as a function of sample number, at a low magnetic field of 
$g\mu_BB_0/J_0 =0.1$. The two lowest energy levels $E_0$ and $E_1$ fluctuate in sync, 
so that the gap $\Delta_{01}$ is free from the randomness of $J_i$ 
[shown in Fig.~\ref{Fig:Odd_RJ_Gap}(b)]. The energies $E_2$ and $E_3$ of
$|2\rangle$ and $|3\rangle$ are also in sync, as shown in Fig.~\ref{Fig:Odd_RJ_Gap}~(a). 
However, the gap $\Delta_{12}$, which is a measure of the isolation of the ground doublet from 
the excited states, does fluctuate [as shown in Fig.~\ref{Fig:Odd_RJ_Gap}(b)], because $E_1$ 
and $E_2$ have different dependence on the exchange coupling. In other words, while the ground 
state splitting $\Delta_{01}$ of this odd-size bus is robust against the randomness in exchange 
coupling, the ground-excited-state gap $\Delta_{12}$ is affected by the randomness. In 
Figs.~\ref{Fig:Odd_RJ_Gap}~(c) and (d) the two gaps, $\Delta_{01}$ and $\Delta_{12}$, are 
plotted as a function of the uniform magnetic field $B_0$ applied on the bus. At low fields, 
the ground state gap $\Delta_{01}$ increases linearly and without broadening as the magnetic 
field increases, until $g\mu_BB_0/J_0 \sim 0.35$. It starts to be influenced by the exchange 
randomness above $g\mu_BB_0/J_0 \sim 0.35$, which corresponds to the crossing between levels 
$\ket{2}$ and $\ket{3}$, as shown in Fig.~\ref{Fig:Odd_Gap} in Appendix~\ref{app:odd_bus}. 
Beyond this crossing point, states $|1\rangle$ and $|0\rangle$ are not the time-reversal of 
each other anymore due to the level crossings with higher excited states. Recall that for 
a spin chain 
to act as a spin bus, we need the ground state doublet to be well separated from the excited 
states, or $\Delta_{12} \gg \Delta_{01}$. Panels (c) and (d) of Fig.~\ref{Fig:Odd_RJ_Gap} 
indicate that this condition is satisfied when the bus ground doublet is energetically 
separated from excited states ($g\mu_BB_0/J_0 \lesssim 0.2$), and acts as an effective 
spin-1/2 system with a constant magnetic moment.
 
\begin{figure}[htbp]
\includegraphics[scale=1.0]{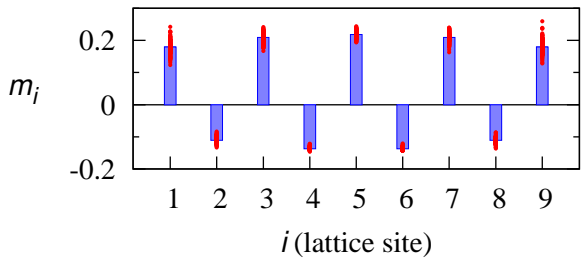}
\caption{(Color online) Dimensionless local magnetic moments 
$m_i=\bra{0;\tfrac{1}{2}} \sigma_{iz}\ket{0;\tfrac{1}{2}}$ are plotted 
(red dots) for all lattice sites on an odd-size spin bus with $N=9$
for an ensemble of $N_S=100$ random samples, with $\sigma_J/J_0=0.01$ and $B_0=0$.
The height of the 
blue bars represents the local magnetic moment $m_i^{(0)}$ for uniform exchange couplings 
($\delta J=0$).
The spread of the red dots at each bus node illustrates the variations in local magnetic 
moments due to the random exchange couplings, with larger variations apparent at the end sites.}
\label{Fig:local_moment}
\end{figure}

Although $[H_C,S_z] = 0$ dictates that the dimensionless total magnetic moment 
$\sum_im_i = \pm 1$ of the ground state is still a good quantum number despite random exchange 
couplings, the local magnetic moment 
$m_i$ does fluctuate around $m_i^{(0)}$, as shown in Fig~\ref{Fig:local_moment}.  Consequently,
the first-order effective coupling $J_{\alpha,i}^{(1)}$ between the qubit and the bus, 
given by Eq.~(\ref{Eq:coupling_J1}), is affected by the randomness in the bus exchange coupling, 
and has to be calibrated individually.
 
\begin{figure}[htbp]
\includegraphics[scale=1.0]{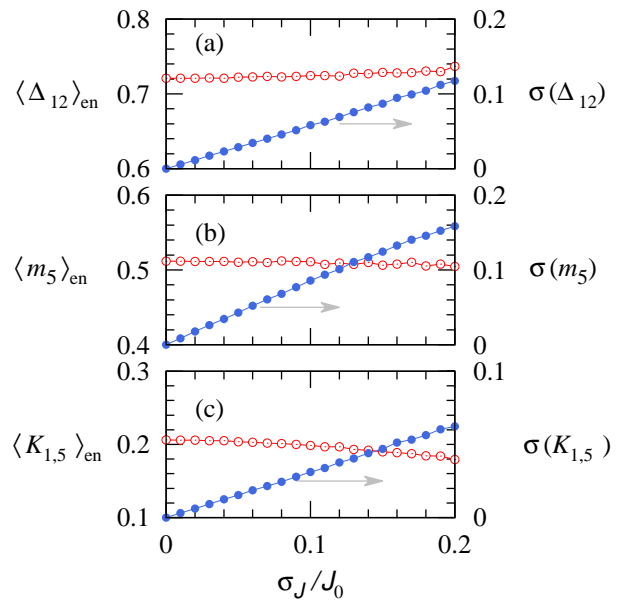}
\caption{(Color online) For odd-size chain with $N=5$, the ensemble averages of (a) the gap 
$\langle\Delta_{01}\rangle_{\rm en}$ in units of $J_0$, (b) the dimensionless local magnetic 
moment $\langle m_5\rangle_{\rm en}$ at the end of the chain, and (c) the normalized 
second-order effective coupling 
$\langle K_{i,j}\rangle_{\rm en}=\langle J^{(2)}_{1,5}\rangle_{\rm en} J_0/(J_{A,1}J_{B,5})$ 
are plotted (open red circle) as a function of the standard deviation $\sigma_J$ of 
$\delta J_i$.  The blue filled circles in each panel indicate the standard deviations of 
each data point, which is obtained by averaging over 2000 random configurations. Here $B_0 = 0$.}
\label{Fig:odd_random_J}
\end{figure}

The effective qubit-qubit coupling $J^{(2)}_{i,j}$ is also affected by the randomness of the 
intra-bus exchange couplings $J_i$. As indicated in Eq.~(\ref{Eq:J_RKKY}), $J^{(2)}_{i,j}$ is 
determined by the full bus spectrum, including both the energy levels and the excited states. 
Figures \ref{Fig:Odd_RJ_Gap} and \ref{Fig:local_moment} show that the random exchange couplings 
$J_i$ in general affect the energy gaps from the ground state, $E_n - E_0$, as well as the bus 
eigenstates $\ket{n}$. Thus we expect that $J^{(2)}_{i,j}$ should be sensitive to the randomness 
in $J_i$. Figure \ref{Fig:odd_random_J} shows how the ensemble averages of the gap, 
$\langle\Delta_{01}\rangle_{\rm en}$, the local magnetic moment, $\langle m_5\rangle_{\rm en}$ 
(which is the normalized first-order qubit-bus coupling $J^{(1)}_{A,5}/J_{A,5}$), and 
the normalized second-order effective coupling, $\langle K_{i,j}\rangle_{\rm en} 
= \langle J^{(2)}_{i,j}\rangle_{\rm en}\,{J_0}/{(J_{A,1}J_{B,5})}$ depend on the fluctuations 
of the exchange coupling, represented by the standard deviation $\sigma_J$, over 
a 5-node bus. The blue filled circles 
in Fig.~\ref{Fig:odd_random_J} show how the fluctuations in these quantities depend on 
the randomness in the exchange couplings.  For example, when 
$\sigma_J/J_0 = 0.1$, $\sigma(m_5) \sim \sigma(K_{1,5}) \sim \frac{1}{6}$, which indicates 
that the effective qubit-bus coupling and the effective qubit-qubit coupling have similar 
sensitivities to the random variations in the intra-bus exchange couplings. 

Furthermore, both 
the local magnetic moment (thus the qubit-bus coupling) and the effective qubit coupling are linear 
functions of $\sigma_J$, with their slopes depending on the size of the bus. These slopes are 
indicators of sensitivity of $J^{(1)}$ and $J^{(2)}$ to the exchange variations, and can be 
used in evaluating decoherence in such a spin bus architecture. For example, background charge 
fluctuations can affect exchange couplings between neighboring nodes of a spin bus. As a result, 
the effective qubit-qubit exchange coupling becomes a time-dependent random variable, which leads 
to two-qubit dephasing.~\cite{HuPRL06} The relevant correlation function that determines 
the dephasing is $\langle J^{(2)}(t) J^{(2)}(0)\rangle$, and is given approximately by
$\left[\sigma(J^{(2)})/\sigma_J \right]^2 {\langle J_i(t) J_i(0)\rangle}$~\cite{HuPRL06}. 
The latter correlation function, $\langle J_i(t) J_i(0)\rangle$, represents fluctuations in 
the individual inter-node 
exchange couplings along the bus, whose dynamics is determined by the environmental charge noise.

In summary, even when intra-bus exchange couplings of an odd-size spin chain have random but 
static variations, the chain can still act as a spin bus, with a ground state doublet that is 
well separated from the excited states, and acts as an effective spin-1/2 system 
with a constant magnetic moment.
However, the effective qubit-bus couplings and the mediated qubit-qubit couplings are affected 
by the randomness in exchange, with their fluctuations linearly proportional to the randomness 
in exchange.  Calibration would thus be needed for accurate qubit operations. The results
here also have implications for spin-bus related decoherence.  In essence, the strong exchange 
couplings allow a spin bus to process quantum information across a large distance, but also 
make the qubit-bus system susceptible to charge noise via both $J^{(1)}_{i,\alpha}$ and
$J^{(2)}_{i,j}$.

\subsection{Effects of Random Magnetic Fields in Odd-size Buses}
\label{subsec:Odd_RandomB}

Spin qubits are generally susceptible to magnetic noise, and the spin bus is no exception. Here we 
examine how random but static external magnetic fields affect the properties of an odd-size spin 
bus. Our results should also be a useful indicator of the sensitivity of a spin bus to temporal
magnetic noise, as we will discuss later in the section.  For this calculation we assume that 
the exchange couplings $J_i$ are uniform, and focus on the magnetic randomness.

The local magnetic fields, Eq.~(\ref{random_B}), are
\begin{align}
B_i = B_0 + \delta B_i\,,
\end{align}
where the random field $\delta B_i$ is sampled from a normal distribution with standard deviation 
$\sigma_B$.  As in the previous subsection, we consider an ensemble of $N_S$ spin chains, so that 
the ensemble average of local magnetic field $B_i$ is $\mean{B_i}_{\rm en} = B_0$ and 
$\mean{\delta B_i}_{\rm en} = 0$ in the limit of large $N_S$. Such a random distribution of local 
magnetic field could originate from quasi-static nuclear hyperfine fields, or local paramagnetic 
centers in a semiconductor.

\begin{figure}[htbp]
\includegraphics[scale=1.0]{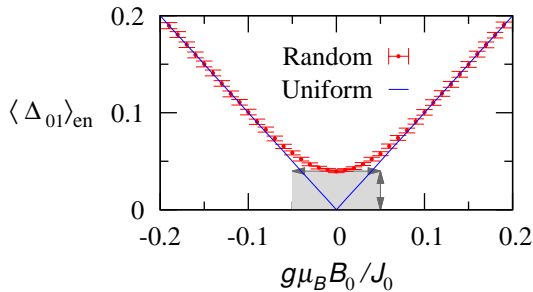}
\caption{(Color online) Ensemble average of the gap $\mean{\Delta_{01}}_{\rm en}$ in units of
$J_0$ between 
the two lowest states of a single spin as a function of the external magnetic field $B_0$. 
The height and the width of the rectangle, touching the bottom of the curve, indicates 
$g\mu_B\sigma_B/J_0$ and $\mean{\Delta_{01}}_{\rm en}/J_0$ at $B_0=0$, respectively.} 
\label{Fig:odd_randomB_gap}
\end{figure}

As a benchmark, we first recall how an odd-size spin chain behaves in a constant uniform 
magnetic field. When $B_0 = 0$, the odd-size chain has two doubly degenerate ground states 
with the total magnetic moment $S_z = \pm 1/2$. A finite $B_0$ splits these two states like 
a single spin-$1/2$ particle. There are two important features that determine the behavior 
of the ground doublet in a magnetic field: the energy splitting $\Delta_{01}$, and the ground 
state spin orientation. In a uniform field the latter depends only on the $g$-factor of the 
material, while in a random field it also depends on the local field configuration.

We first examine how the random external magnetic field affects $\Delta_{01}$, the splitting of 
the ground state doublet of the odd-size bus. This dependence is a good indicator of whether 
the ground state doublet of an odd-size spin bus is a robust effective spin-1/2. Note that the
commutation relation $[H_C,S_z] = 0$ holds even in a random magnetic field, so the two lowest 
states have $S_z = \pm 1/2$ (recall that the inter-node coupling is anti-ferromagnetic). 
Figure~\ref{Fig:odd_randomB_gap} shows that the randomness in the magnetic field induces a finite
average gap at zero field. This gap opens at $B_0 = 0$ even though 
$\mean{\delta B_i}_{\rm en} = 0$. This non-vanishing average gap between the two lowest states 
is a consequence of the statistical behavior of the random field, and can be understood using 
a model of $N_S$ single spins in a Gaussian ensemble of random external magnetic fields 
$\delta B$ with $B_0 = 0$. The Hamiltonian of a single spin is
\begin{align}
H = -g\mu_B\,\delta B\,\frac{\sigma_z}{2}\,.
\end{align}
The energy splitting of each spin is given by $\Delta_{01} = g\mu_B|\delta B|$, which is 
{\it always} positive. It is thus not a surprise that the ensemble average of the gaps, 
$\mean{\Delta_{01}}_{\rm en} = g\mu_B\mean{|\delta B|}_{\rm en}$, is nonzero, even though 
$\mean{\delta B}_{\rm en} = 0$---the ground state changes according to the field configuration. 

To make this argument more rigorous, recall that for a normal distribution, the odd central 
absolute moments of a random variable 
$X$ with a mean of $\mu$ are given by
\begin{align}
{\mathbb E}\left(|X-\mu|^p\right)=\frac{\sigma^p (1-p)!!}{\sqrt{2\pi}}\,,
\label{odd_expect}
\end{align}
where $!!$ denotes the double factorial. Applying Eq.~(\ref{odd_expect}) to $\Delta_{01}$, 
we get
\begin{align}
\mean{\Delta_{01}}_{\rm en} = g\mu_B\frac{\sigma_B}{\sqrt{2\pi}}\,,
\end{align}
where $\sigma_B$ is the standard deviation of $\delta B$. While in our case the magnetic moment 
of the ground state is distributed throughout the whole spin chain, its splitting is still 
{\it mainly} due to the Zeeman splitting of a single Bohr magneton, so that the single-spin 
argument provided here is still applicable.  When $B_0 \neq 0$, the average gap can be 
qualitatively expressed as 
$\mean{\Delta_{01}}_{\rm en} \propto g\mu_B \sqrt{\mean{(B_0 + \delta B)^2}} 
\sim g\mu_B \sqrt{B_0^2 + \sigma_B^2}$, which is a hyperbola that saturates at $B_0 = 0$ to $g
\mu_B \sigma_B$ and approaches $g \mu_B B_0$ at large $B_0$. The difference in 
the proportionality constant originates from the difference between $\mean{|f|}$ 
and $\sqrt{\mean{f^2}}$.

\begin{figure}[htbp]
\includegraphics[scale=1.00]{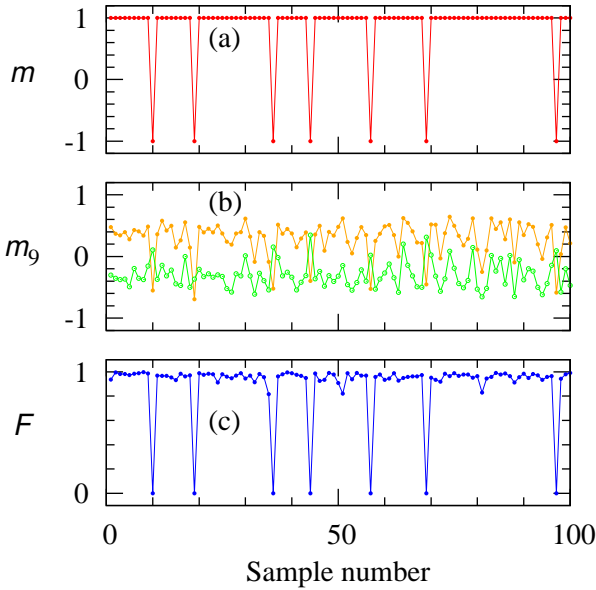}
\caption{(Color online) (a) Total magnetic moment $m$ of the ground state $\ket{0}$, (b) local 
magnetic moment $m_9$ for the ground state $\ket{0}$ (orange solid circle) and for the first 
excited state $\ket{1}$ (green open circle), and (c) the fidelity $F$ of the ground state in
the presence of random magnetic fields with respect to that in the uniform field as a function 
of the sample number. Here the size of spin buses is $N=9$, $g\mu_BB_0/J_0 = 0.1$, 
and $g\mu_B\sigma_B/J_0 = 0.065$.} 
\label{Fig:rand_B_moment}
\end{figure}

\begin{figure}[htbp]
\includegraphics[scale=1.00]{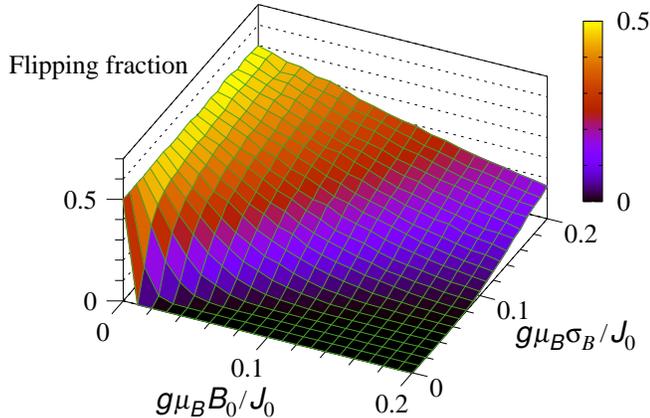}
\caption{(Color online) The fraction of flipped ground states for $N_s= 10000$ random 
realizations of odd-size chains with $N=5$, as a function of $B_0$ and $\sigma_B$.}
\label{Fig:Odd_RB_sigma}
\end{figure}

Now we address the spin orientation of the odd-size spin bus in the ground state under 
random magnetic fields. The magnetic moment of 
a single spin (with $g=2$) in the ground state is anti-parallel to the external field.
As in the single spin case, the total magnetic moment of the odd-size spin bus 
in the ground state is always anti-parallel to a {\it small uniform} external magnetic 
field. As shown in Fig.~\ref{Fig:local_moment}, the local magnetic moments align 
antiferromagnetically (alternation between anti-parallel and parallel alignments).
Even in a random 
external 
magnetic field in the $z$-direction, the two lowest states of an odd-size 
spin bus are characterized by $S_z=1/2$ or $S_z=-1/2$, so one might guess
the Zeeman energy would determine the spin orientation. 
However, this is not the case.

Figure~\ref{Fig:rand_B_moment} shows the total magnetic moment of the ground state 
$\ket{0}$ in an ensemble of random magnetic fields for the case when $N = 9$ and 
$\sigma_B/B_0=0.65$. 
As illustrated in panels (a) and (c), seven samples, {\#10}, {\#19}, {\#36}, {\#44}, 
{\#57}, {\#69}, and  {\#97}, among 100 samples have negative total magnetic moment 
$m=-1$, so that their ground state is $\ket{n=0;S_z=-\tfrac{1}{2}}$ instead of
$\ket{n=0;S_z=\tfrac{1}{2}}$.
Panel (b) of Fig.~\ref{Fig:rand_B_moment} shows a strongly random local magnetic moment 
on the 9-node spin bus, with the seven flipped states having reversed local magnetic moments. 
The probability for either $\ket{n=0;S_z=-\tfrac{1}{2}}$ or $\ket{n=0;S_z=\tfrac{1}{2}}$ to 
be the ground state depends on the ratio of $B_0$ and $\sigma_B$, as illustrated in 
Fig.~\ref{Fig:Odd_RB_sigma}. As expected, when $B_0 = 0$, the probability is 50\%. A strong 
uniform field $B_0$ (compared to $\sigma_B$) suppresses the flipping fraction and stabilizes 
one of the states as the predominant ground state.

A close inspection of the data presented in Fig.~\ref{Fig:rand_B_moment} reveals that Zeeman 
splitting of the local nodes of a spin bus does not tell the whole story of $\Delta_{10}$, 
so that the single-spin model has its limitations.  By definition, the energy gap is
\begin{eqnarray}
\label{Eq:Energy_Gap}
\Delta_{10} = E_1 - E_0 &=& \bra{1}H_J\ket{1}  - \bra{0}H_J\ket{0} \nonumber\\
     &&- \mu_B\sum_i B_i (\,m_i^1 - m_i^0\,) \nonumber \\
 &=& E_J^{1} - E_J^{0} + E_Z^{1} - E_Z^{0} \,.
\end{eqnarray}
Here $H_J$ and $H_Z$ represent the exchange and Zeeman components of $H_C$ in 
Eq.~(\ref{Hamil_random}), $E_Z^{k} = \bra{k}H_Z\ket{k} = - \mu_B \sum_i B_i m_i^k$ and 
$E_J^{k} = \bra{k}H_J\ket{k}$ denote the Zeeman and exchange contributions to the energy 
of state $\ket{k}$, with $k=0,1$, $m_i^k \equiv \bra{k}\sigma_{iz}\ket{k}$ is the local 
magnetic moment at bus node $i$, and we have taken $g = 2$ for simplicity. When the external 
magnetic field is uniform, $B_i=B_0$, the exchange contribution to the energy of $\ket{0}$ 
and $\ket{1}$ in Eq.~(\ref{Eq:Energy_Gap}) are the same, so that $E_J^{1} - E_J^{0} = 0$, 
and $m_i^0$ are equal in magnitude and opposite in sign to $m_i^1$. Thus the gap between 
the two states is given completely by the Zeeman splitting: 
$E_1 -E_0 = 2\mu_BB\sum_i m_i^0 = 2\mu_BB m^0$.  With $g=2$, the net spin of the ground 
state $\ket{0}$ is parallel to the uniform magnetic field ${\bf B}$ like a single spin, 
so that $\ket{0;S_z=1/2}$ is the ground state when $B$ is positive.

\begin{figure}[htbp]
\includegraphics[scale=1.0]{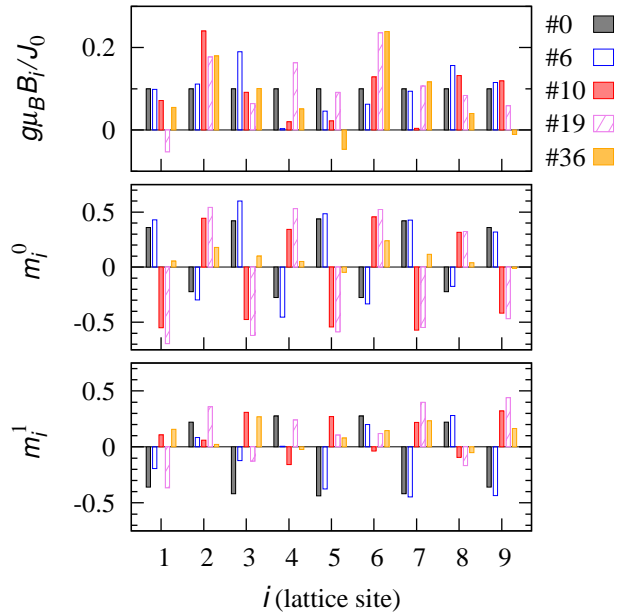}
\caption{(Color online) Local magnetic moments of the ground and first excited states 
for several random field configurations in a 9-node spin bus.
Here sample \#0 corresponds to the case of a uniform applied field. 
The others correspond to the samples in Fig.~\ref{Fig:rand_B_moment}. 
The total magnetic moment of the ground states of 
samples \#0 and \#6 is $m^0=1$, i.e., $S_z = 1/2$. On the other hand, samples \#10, \#19, 
and \#36 have the spin-flipped ground states, $m^0=-1$, i.e., $S_z =-1/2$.}
\label{Fig:N9RB_flip}
\end{figure}

Under a random magnetic field $B_i$, but with $B_0 \gg \sigma_B$, the exchange contribution 
to the gap in Eq.~(\ref{Eq:Energy_Gap}) is still negligible, and the ground state spin is 
parallel to $B_0$.  However, if $B_0$ is of the order of $\sigma_B$ or smaller, the ground 
state total spin orientation may be anti-parallel rather than parallel to $B_0$ . 
For these configurations with the ``flipped'' ground state, the reason for the flipping is 
varied. Consider the samples (from the ensemble of 9-node buses presented in 
Fig.~\ref{Fig:rand_B_moment}) shown in Fig.~\ref{Fig:N9RB_flip} and Table~\ref{tab2}. 
Here sample {\#0} refers to the spin bus in a uniform magnetic field. Samples {\#6}, {\#10},
{\#19} and {\#36} are for the bus in different random magnetic field configurations, and 
the later three samples have a flipped ground state, with $m^0=-1$. Figure~\ref{Fig:N9RB_flip} 
shows that in a random magnetic field, the local magnetic moments for the two lowest states 
are generally not equal in magnitude, $|m_i^0|\ne |m_i^1|$, so that the two states $\ket{0}$ 
and $\ket{1}$ are no-longer spin-flipped image of each other, even though 
$\sum_i m_i^k = \pm 1$. The most dramatic examples are those samples with a flipped ground 
state, where the ground and first excited states are far from the classical 
anti-ferromagnetic spin configurations. This also implies that in general the Zeeman energy 
$E_Z^0 \neq -E_Z^1$. Samples {\#6} and {\#10} demonstrate that even if all the magnetic 
fields $B_i$ are positive, the ground state could still be either $S_z=1/2$ or $S_z = -1/2$. 
While in most cases the energy gap between the two lowest states in Eq.~(\ref{Eq:Energy_Gap})
is due to the Zeeman term, for samples {\#10} and {\#19} the Zeeman energies $E_Z^{0}$ and 
$E_Z^{1}$ are almost equal, so that there is no Zeeman gap. The energy gaps for these two 
samples are determined by the exchange contribution in Eq.~(\ref{Eq:Energy_Gap}), and are 
one or two orders of magnitude smaller than the usual Zeeman gap. For sample {\#36}, 
the total energy gap is dominated by the Zeeman contribution (with significant contribution 
also coming from the exchange interaction), although the configuration of the random 
magnetic field is such that the ground state is flipped. In short, when 
$B_0 \lesssim \sigma_B$, the local and total magnetic moments of the ground state are 
sensitively dependent on the distribution of the random magnetic field.

\begin{table}[htbp]
\begin{tabular}{c|cccc}
\hline
Sample No. & $E_1-E_0$  & $E_Z^1 -E_Z^0$  & $E_J^1 -E^0_J$  & $\quad|\Delta E_J/\Delta E_Z|$\\
\hline
\#0  &0.2     & 0.2            & 0.0           & 0     \%\\[3pt]
\#6  &0.27083 &\quad\, 0.25840 &\quad 0.01244  & 4.8   \%\\[3pt]
\#10 &0.00403 &\quad  -0.00840 &\quad 0.01243  & 148.0 \%\\[3pt]
\#19 &0.06222 &\quad  -0.01582 &\quad 0.07804  & 493.3 \%\\[3pt]
\#36 &0.06989 &\quad\, 0.04306 &\quad 0.02683  & 62.3  \%\\
\hline
\end{tabular}
\caption{(Color online) Energy gap $E_1 -E_0$ between the bus states $\ket{0}$ and 
$\ket{1}$, its Zeeman contribution $E_Z^1-E_Z^0$, the exchange contribution 
$E_J^1-E_J^0$, and the absolute ratio of these contributions for samples {\#0}, {\#6}, 
{\#10}, {\#19}, and {\#36} in Fig.~\ref{Fig:rand_B_moment}. Here the energy is measured 
in units of $J_0$.}
\label{tab2}
\end{table}

Our results so far indicate that random magnetic fields can seriously undermine the 
capabilities of a spin bus, by altering the local magnetic moments (and thus the effective 
qubit-bus coupling $J^{(1)}$) and the bus ground state.  Fortunately, in general $\sigma_B$ 
is relatively small. For example, if the random field is due to hyperfine interaction 
in GaAs, $\sigma_B \sim 2$ mT for a 100 nm quantum dot, so that a $B_0 > 20$ mT should be 
more than enough to overcome the effect of the random field. Furthermore, the above 
discussion is applicable to the regime where the magnetic energy scales are not much 
smaller than the exchange energy scales (e.g., for Fig.~\ref{Fig:rand_B_moment} and 
Table~\ref{tab2}, $\mu_B\sigma_B/J_0 = 0.0650$). If $J_0 \gg B_0$ and $\sigma_B$, 
the structures of the ground and first excited states of the spin chain should be 
determined by the anti-ferromagnetic coupling and are less susceptible to the small 
magnetic field or its fluctuations. In this case the ground state spin orientation 
would be mostly determined by the Zeeman contribution to the energies of the two states, 
and we would recover the simple single-spin physical picture.

\begin{figure}[htbp]
\includegraphics[scale=1.0]{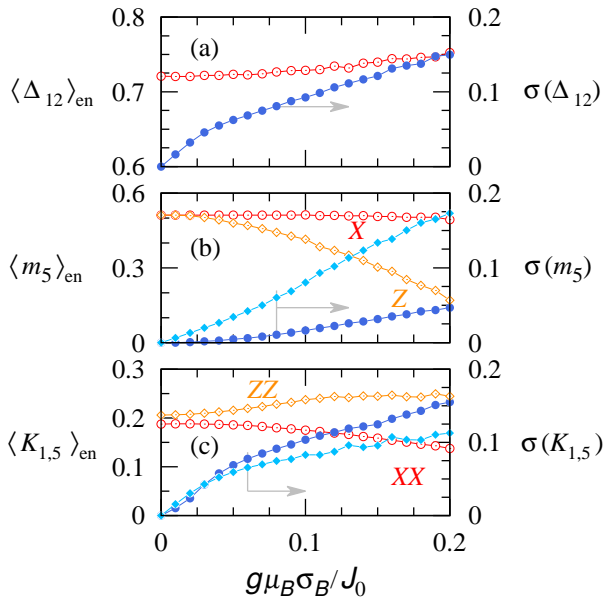}
\caption{(Color online) For an odd-size spin bus with $N=5$, the ensemble averages of 
(a) the gap $\langle\Delta_{12}\rangle_{\rm en}$ in units of $J_0$, (b) the local magnetic 
moment $\langle m_5\rangle_{\rm en}$ on one end of the chain, and (c) the normalized 
second-order effective coupling $\langle K_{1,5}\rangle_{\rm en}$ are plotted as a function 
of the standard deviation $\sigma_B$ of $\delta B_i$. In (b), the labels ``X'' 
(symbol: red open circle) and ``Z'' (symbol: orange open diamond) stand for the $x$- 
and $z$- components of the first-order effective coupling, respectively. In (c) the labels 
``XX'' (symbol: red open circle) and ``ZZ'' (symbol: orange open diamond) refer to 
the $xx$- and $zz$-component of the second-order effective coupling, respectively. 
Here $\mu_BB_0/J_0 = 0.05$.  In panels (b) and (c), the solid diamonds and solid circles 
represent standard deviations for the data labeled by the open diamonds and open circles, 
respectively.}
\label{Fig:odd_random_B}
\end{figure}

The fluctuations in the local magnetic moments of the spin bus due to the random external 
magnetic field lead directly to fluctuations in the effective qubit-bus couplings 
$J^{(1)}_{i,\alpha}$ [see Eq.~(\ref{Eq:coupling_J1})], though this fluctuation is 
suppressed if $B_0 \gg \sigma_B$, as illustrated by the few data points for small 
$\sigma_B/B_0$ in Panel (b) of Fig.~\ref{Fig:odd_random_B}.
In addition, under a random magnetic field, the effective coupling $J_{\alpha,i}^{(1)}$, 
given by Eq.~(\ref{Eq:coupling_J1}), becomes anisotropic.  This is in addition to 
the anisotropy induced by a finite $B_0$.\cite{Shim12} In general, anisotropy occurs 
when
\begin{subequations}
\begin{align}
\bra{0} \sigma_{ix}\ket{1} 
&\ne \frac{1}{2}\Bigl[\,\bra{0} \sigma_{iz}\ket{0} - \bra{1} \sigma_{iz}\ket{1}\,\Bigr]\,,
\end{align}
and
\begin{align}
\sum_{i=1}^{N}& \bra{0} \sigma_{ix}\ket{1} \ne 1\,.
\end{align}
\end{subequations}
The anisotropy appears in both $J^{(1)}$ and $J^{(2)}$~\cite{Shim11}, as illustrated 
in Panels (b) and (c) of Fig.~\ref{Fig:odd_random_B}. Based on the standard deviation data 
presented in panel (b), we also observe that while the transverse component of the local 
magnetic moment is reasonably robust against the randomness in the magnetic field, the 
longitudinal component is not.  On the other hand, panel (c) shows that both the longitudinal 
and transverse components of the effective qubit-qubit coupling $J^{(2)}$ have a linear 
dependence on the field randomness for small $\sigma_B$, changing to a different slope as 
$\sigma_B$ becomes larger than $B_0$.  Both of these observations illustrate the fact that
for an odd-size spin bus to function properly, magnetic field randomness in the system 
needs to be minimized.

As explained in Sec.~\ref{sec:review}, the effective qubit-qubit coupling $J^{(2)}$ 
given by Eq.~(\ref{Eq:J_RKKY}) is well defined when the degenerate or nearly degenerate 
ground states of the spin bus are spin-flipped states of each other. This is the case for
an odd-size spin bus near zero external magnetic field. At certain finite external fields, 
the ground states of the spin bus would be close to be degenerate again. However, in those 
regimes Eq.~(\ref{Eq:J_RKKY}) is generally not applicable and should be modified.~\cite{Shim12}

In summary, the effects of the random magnetic field in the $z$-direction on an odd-size 
bus are as follows.  First, the degeneracy in the ground states of an odd-size bus is lifted.
Second, although the two lowest states have either $S_z=1/2$ or $S_z=-1/2$, the spin 
orientation of the ground state is not solely determined by the Zeeman energy. 
Third, the random magnetic fields make the first-order and the second-order effective 
couplings anisotropic.

\subsection{Effects of Random Exchange Couplings in Even-size Buses}

In this subsection, we investigate how an even-size spin bus is affected by random exchange 
couplings $J_i$.  Recall that the effective Hamiltonian for an even-size bus coupled to two 
qubits in zero magnetic field takes the form of 
$H_{\rm eff}^{(2)} = J^{(2)}_{i,j}\,{\bf S}_A\cdot{\bf S}_B$, where $J^{(2)}_{i,j}$ 
is given by Eq.~(\ref{Eq:J_RKKY}).  The various terms in this equation are the bare 
qubit-bus couplings $J_{A,i}$ and $J_{B,j}$, the bus energy gaps $\Delta_{0n} = E_n - E_0$, 
and the transition matrix elements $\bra{0}\sigma_{i\mu}\ket{n}$. Here we examine the effects 
of random intra-bus exchange couplings on these different terms, with a particular focus
on the gap $\Delta_{01}$ between the ground and the first excited state, which is an indicator 
of how well the nondegenerate ground state is isolated from the excited states, and figures 
prominently in the expression of $J^{(2)}$.

\begin{figure}[htbp]
\includegraphics[scale=1.0]{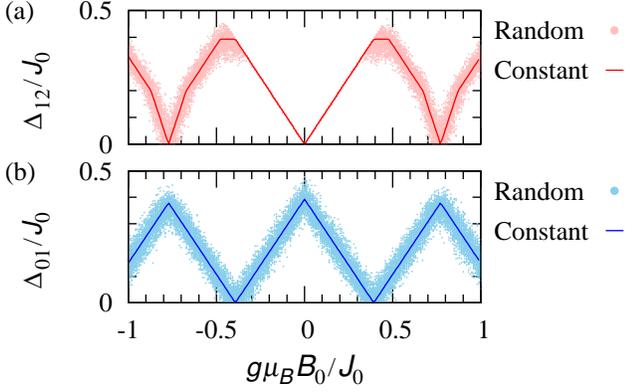}
\caption{(Color online) For $N_s=100$ even-size chains with random exchange couplings, 
two lowest energy gaps, (a) $\Delta_{12} = E_2-E_1$ and (b) 
$\Delta_{01} =E_1-E_0$, are plotted as functions of $B_0$. The size of 
each chain is $N=8$ and the standard deviation $\sigma_J/J_0$ of $\delta J_i$ is 0.025.}
\label{Fig:Even_RJ_Gap}
\end{figure}

An even-size bus can be thought of as an odd-size bus plus an extra spin-1/2 node. 
Thus the lowest four states of an even-size bus is a singlet and a triplet. A uniform 
magnetic field would split the triplet, but would not affect the singlet ground state. 
With random exchange couplings, similar to the case of an odd-size bus, we still have 
$[H,S_z] = 0$, so that the triplet splitting is given by Zeeman splitting.  
The random exchange couplings do cause the energy levels and the eigenstates to vary 
in general.  For example, Fig.~\ref{Fig:Even_RJ_Gap} plots the two energy gaps, 
$\Delta_{01}$ and $\Delta_{12}$, as 
a function of $B_0$ for 100 samples of even-size chains with random exchange couplings. 
While the ground state remains nondegenerate, the gap $\Delta_{01}$ is now distributed between 
0.3 and $0.5\, J_0$ at $B_0=0$. Around $B_0=0$, the gap $\Delta_{12}$ and the next gap 
$\Delta_{23}= E_3-E_2$ (not plotted) are robust against the random exchange 
couplings $J_i$, reflecting the fact that these two gaps correspond to the Zeeman splittings 
of the triplet bus states.  This is similar to the $\Delta_{01}$ gap of an odd-size bus 
near zero field, which corresponds to the Zeeman splitting of the spin-1/2 doublet 
ground state, as shown in Fig.~\ref{Fig:Odd_RJ_Gap}.

\begin{figure}[htbp]
\includegraphics[scale=1.0]{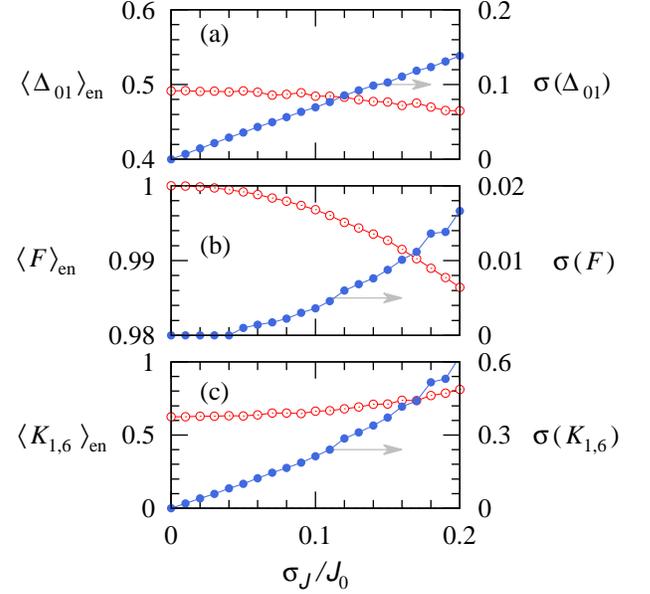}
\caption{(Color online) For an even-size bus with $N=6$, (a) the ensemble averages of 
the ground energy gap $\Delta_{01}$, (b) the fidelity $F$ with respect to the uniform 
ground state, and (c) the normalized effective qubit-qubit coupling 
$K_{1,6} = J^{(2)}_{1,6} J_0/J_{1,A}J_{6,B}$ are plotted as functions of the fluctuation 
$\sigma_J$ of the random exchange coupling (all data are represented by red open circles). 
Here $B_0 = 0$ and the sample size is $N_s=100$. The blue solid circles represent 
the standard deviation of $\Delta_{01}$, $F$, and $K_{1,6}$, respectively.}
\label{Fig:Even_RJ_Coupling}
\end{figure}

The most important effect of the randomness in the exchange couplings for bus
operations is on the low lying energy gaps such as $\Delta_{01}$.  As shown in panel 
(b) of Fig.~\ref{Fig:Even_RJ_Coupling}, the ground state itself is quite robust against 
the random exchange coupling in terms of state fidelity, and has zero 
local magnetic moments as well as zero total magnetic moment.
This is in contrast with the effect of random magnetic fields on the even-size bus
as shown in subsection~\ref{subsec:even_randomB}, where the local magnetic moments 
become non-zero. On the other hand, while 
the average value of the $\Delta_{01}$ gap is only weakly dependent on $\sigma_J$, 
the fluctuations in $\Delta_{01}$ depend linearly on $\sigma_J$, as shown in panel (a) 
of Fig.~\ref{Fig:Even_RJ_Coupling}.  Consequently, the fluctuations of the bus-mediated 
qubit-qubit coupling also has a linear dependence on the standard deviation $\sigma_{J}$ 
of $\delta J_i$, as shown in panel (c) of Fig.~\ref{Fig:Even_RJ_Coupling}. The slope 
$\sigma_{J^{(2)}}/\sigma_J$ is quite large here, reflecting a sensitive dependence of the
singlet-triplet splitting of the spin bus on the intra-bus exchange couplings.  
While calibration should be able to largely suppress the effects of any static randomness 
of the exchange coupling, the sensitivity to randomness in the local exchange, as indicated 
by the large $\sigma_{J^{(2)}}/\sigma_J$, dictates that effects of environmental charge 
noise on the inter-node exchange coupling $J_i$ need to be minimized.

\subsection{Effects of Random Magnetic Fields in Even-size Buses}
\label{subsec:even_randomB}

In this subsection, we study how an even-size bus is affected by an external magnetic field 
that has random local variations. In a uniform but small external magnetic field,
the ground state of the even-size bus has zero local magnetic moments as well as 
zero total magnetic moment ($S_z=0$). 
Recall that the commutation relation $[H_C,S_z]=0$ is 
still valid even when the bus is subject to random magnetic fields $B_i$ in the $z$-direction. 
If the random magnetic fields are weak, the even-size bus remains in  
the ground state with $S_z=0$, i.e., zero total magnetic moment. However, the local 
magnetic moments $m_i$ become non-zero in contrast with the case of a uniform magnetic field.
Furthermore, the bus excited states generally do have net magnetic moments, 
so that they respond to both local and global 
magnetic fields in terms of their energies and their state composition.

\begin{figure}[htp]
\includegraphics[scale=1.0]{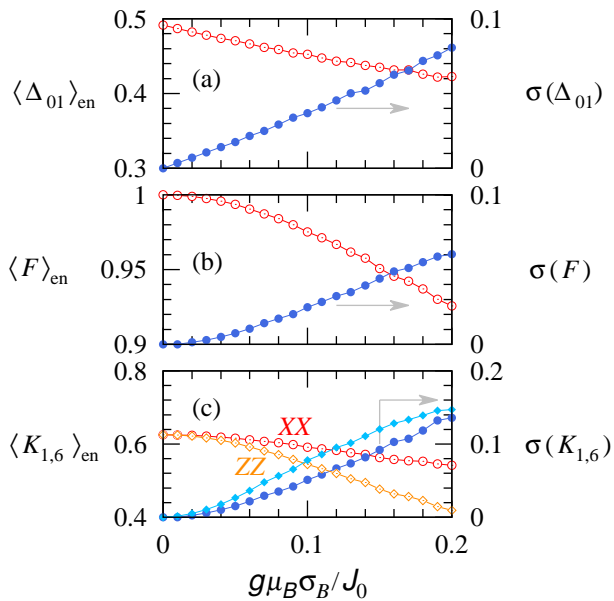}
\caption{(Color online) Effect of random magnetic field for an even-size bus with $N=6$. 
(a) The average ground energy gap $\Delta_{01}$ in units of $J_0$, (b) the average fidelity 
$F$ of the ground states with respect to that without random $B_j$, and (c) the average of 
the normalized second-order effective coupling $K_{1,6}$ are plotted (in open red circles 
or open orange diamonds) as a function of the standard deviation of the random magnetic field 
$\sigma_B$. 
The standard deviations of the averages are given by the blue solid circles or blue 
solid diamonds, respectively.  Here we take $B_0 = 0$.} 
\label{Fig:Even_RB}
\end{figure}

In Fig.~\ref{Fig:Even_RB} we plot the effect of the locally random magnetic field on the gap 
$\Delta_{01}$, the ground state robustness (in terms of the fidelity $F$ with respect to the 
uniform field ground state), and the bus-mediated qubit-qubit interaction $J^{(2)}_{1,6}$. 
The average of gap $\Delta_{01}$ decreases with $\sigma_B$ because for any particular random field 
configuration, one of the polarized triplets has a lower energy compared to the zero field 
case and becomes the first excited state. The linear increase in the standard deviation of 
$\Delta_{01}$ is simply a reflection of the linear nature of the Zeeman splitting. 
In panel (b), the decrease in the fidelity shows that the local magnetic 
moments may fluctuate although their sum, i.e., the total magnetic moment is still zero.
Interestingly, 
in panel (c), the normalized effective coupling $K_{1,6}=J^{(2)}_{1,6} J_0/J_{A,1}J_{B,6}$ and 
its standard deviation both have a super-linear dependence on $\sigma_J$ when it is small, 
making the coupling a robust quantity against field randomness. The qualitative reason for 
this robustness is that the ground state is not coupled to the two polarized triplet states 
of the bus by the random magnetic field along $z$ direction at the lowest order, 
while the gap between the unpolarized triplet state and the ground state only 
depends on the field randomness quadratically. In this calculation we did not consider 
a random transverse field since our focus is on static disorder. For example, for magnetic 
disorder caused by random nuclear polarizations, the transverse polarizations would precess 
around the external field, so that their effect would tend to be suppressed.

In Ref.~\onlinecite{Shim11}, we have shown that a constant external magnetic field makes 
the second-order effective interaction mediated by an even-size bus anisotropic. Here we find 
that local random variations in the magnetic field, $\delta B_i$, also induce anisotropy, 
even if $B_0=0$ and $\langle \delta B_i\rangle_{\rm en} = 0$, as shown in panel (c) of 
Fig.~\ref{Fig:Even_RB}.  This anisotropy is weaker compared to the case of an odd bus in 
a finite $B_0$ as shown in Fig.~\ref{Fig:odd_random_B}, because the average of the local 
field here vanishes ($B_0 = 0$), so that the spatial isotropy is not broken as completely as 
in the case of Fig.~\ref{Fig:odd_random_B}. Note that we would normally expect an even-size 
bus to be operated at zero external magnetic field $B_0=0$, unless the anisotropy of 
the effective coupling is desired.

\section{Conclusions}
\label{sec:summary}

We have performed a comprehensive study of the effects of local randomness in the exchange 
couplings and the external magnetic field on the capabilities of a strongly coupled 
Heisenberg spin bus.

We find that the random exchange couplings preserve the isotropic symmetry (in the qubit-bus 
and qubit-qubit couplings) of the bare Heisenberg coupling.  This symmetry also makes the ground 
Zeeman energy gap of the odd-size bus robust against small fluctuations in the exchange 
couplings. However, randomness in the exchange couplings does cause the eigenenergies and 
the eigenstates to vary, which in turn leads to randomness in the magnitudes of the effective 
couplings (both qubit-bus and qubit-qubit).

An external magnetic field, whether uniform or random, does break the isotropy of the Heisenberg 
spin bus, and leads to anisotropy in the qubit-bus and qubit-qubit effective couplings. 
A locally random magnetic field also lifts the ground state degeneracy of an odd-size bus, even
when the average applied field vanishes.  The local randomness also gives rise to the effect 
that the total magnetic moment of the odd size bus in the ground state may be {\em antiparallel 
to the direction of the applied magnetic fields}, when a single spin in the ground state would 
have been parallel to the magnetic field. Even-size buses are somewhat more robust against local 
random magnetic fields, since their ground state is non-magnetic.

We have performed ensemble calculations for the coupled qubit-bus systems we have considered, 
where the standard deviation of an ensemble averaged quantity (such as the qubit-bus and 
qubit-qubit effective couplings) represents the sensitivity of this quantity to the particular
parameter randomness. Thus our results have clear implications not only for situations where 
static parameter randomness is present, but also for dynamical noise in the exchange coupling 
or the external field.

To give context to this paper, in the Appendices we provide a comprehensive overview of even 
and odd spin chains as quantum data buses. In particular, we derive the first- and second-order 
effective couplings using the projection operator method. We explore the low-energy spectra of
buses coupled to zero, one or two qubits, from which we derive the first- and second-order 
effective Hamiltonians of the qubit-bus system.  We also prove that random exchange couplings 
do not lift the ground state degeneracy of an odd-size bus. Finally, we present a study of 
the scaling properties of the bus, for up to 20 nodes.

\begin{acknowledgments}
This work was supported by the DARPA QuEST program through AFOSR and NSA/LPS through ARO.
\end{acknowledgments}

\appendix
\section{Derivation of the spin bus effective Hamiltonians by projection method}
\label{app:eff_hamil}

In this Appendix we derive the effective low-energy Hamiltonian of the full 
Hamiltonian~(\ref{total_Hamil}). The weak qubit-bus couplings $J_{\alpha,i}/J_0 \ll 1$ are 
used as perturbation parameters. The total Hamiltonian $H = H_0 + H_1$ can be rewritten as 
an unperturbed Hamiltonian $H_0 = H_C + H_Q$ and a perturbation $H_1=H_{QC}$. Since we are 
interested in the low-energy limit, we define a projection operator $P$ onto the subspace 
${\cal H}_0$ spanned by the tensor products $\ket{\Phi_k}$ of the ground state(s) of 
the free bus Hamiltonian $H_C$ and the eigenstates of the free qubit Hamiltonian $H_Q$
\begin{align}
P = \sum_{k\in {\cal H}_0} \ket{\Phi_k}\bra{\Phi_k}\,.
\end{align}
Here the Zeeman energy of $H_Q$ is assumed to be small compared to the gap of $H_C$, 
so that the energy of the subspace ${\cal H}_0$ is equal or nearly equal to the ground 
energy $E_0$ of $H_C$. In this limit, we can set $H_Q = 0$. The finite-field induced 
anisotropy is addressed elsewhere.~\cite{Shim11,Shim12}  The effective Hamiltonian 
acting on the subspace ${\cal H}_0$ is then given by~\cite{Fulde,Cohen}
\begin{align}
H_{\rm eff} =  PHP + PHQ\,\frac{1}{E_0 -QHQ}\,QHP \,,
\end{align}
where $Q= \mathbb{I} -P$ projects onto the sub-Hilbert-space orthogonal to ${\cal H}_0$. 
The effective Hamiltonian to second order in $H_1$ is given by
\begin{align}
H_{\rm eff}^{(2)} =  PHP + PH_1Q\,\frac{1}{E_0 -H_0}\,QH_1P \,. \label{Eff_Hamil}
\end{align}
The derivation of the explicit form of the effective Hamiltonian~(\ref{Eff_Hamil}) requires 
detailed information on $P$, which consists of the structure and spectrum of the ground 
manifold of $H_C$.

The Heisenberg spin chain Hamiltonian $H_C$, given by Eq.~(\ref{chain_Hamil}), is exactly 
{\it albeit only partially} solvable with the Bethe ansatz.~\cite{Bethe} The $z$ component 
$S_z$ of the total spin ${\bf S} \equiv \sum_i \bs_i$ commutes with 
the Hamiltonian~(\ref{chain_Hamil}), $[H_0, S_z] = 0$, so that the energy eigenstates can 
be labeled by $\ket{n;m_z}$ with the energy level $n$ and the magnetic quantum number
$m_z$, i.e., the eigenvalues of $S_z$. However, the general analytic expressions of 
the eigenstates $\ket{n;m_z}$ are not available.  Traditionally, for bulk systems, 
the periodic boundary condition and an even number $N$ ($N\to \infty$ in thermodynamic limit) 
are assumed. For finite size chains, however, {\it the eigenstates are dependent on 
both the boundary condition and the even-odd parity of size $N$}, as shown in 
Figs.~\ref{Fig:Odd_Eng} and~\ref{Fig:Even_Eng}. Thus, the effective Hamiltonians for 
the bus-qubit system are different depending on the parity of the bus.\cite{Oh10} 
Here we give a more detailed description of the derivation of the effective Hamiltonians. 
Note that Ref.~\onlinecite{Hirjibehedin06} demonstrated experimentally the even-odd parity 
effect of a spin chain, by assembling chains of 1 to 10 Mn atoms on a metallic surface 
and measuring the parity dependent tunneling currents.

\subsection{Effective Hamiltonians with an Odd-Size Bus}
\label{app:odd_bus}

An odd-size antiferromagnetic chain has an odd number of spins, so that the ground state 
should have one uncompensated spin. For example, for $N=3$, the classical antiferromagnetic 
spin configurations are ``up-down-up'' or ``down-up-down''. The exact degenerate quantum 
mechanical ground states of the odd-size chain with $N=3$ around $B_0=0$ are given by
\begin{subequations}
\label{3state}
\begin{align}
\ket{0;+\tfrac{1}{2}}_C
&= \frac{1}{\sqrt{6}}\left( \ket{001} -2\ket{010} + \ket{100} \right) \,,\\
\ket{0;-\tfrac{1}{2}}_C
&= \frac{1}{\sqrt{6}}\left( \ket{110} -2\ket{101} + \ket{011} \right) \,,
\end{align}
\end{subequations}
where $\ket{0}$ and $\ket{1}$ on the right hand side represent the spin up and down states 
of a single spin, respectively. One can see that the basis states corresponding to classical 
configurations, $\ket{010}$ or $\ket{101}$, are most probable, although quantum corrections 
are already sizable. For longer chains, the amplitude of the classical antiferromagnetic 
configuration continues to decrease, while quantum fluctuation contributions increase. 
Although a three-node chain is small, the analytic solution can serve as a starting point 
for understanding how an odd-size chain acts as an effective single spin.

\begin{figure}[ht]
\includegraphics[scale=1.0]{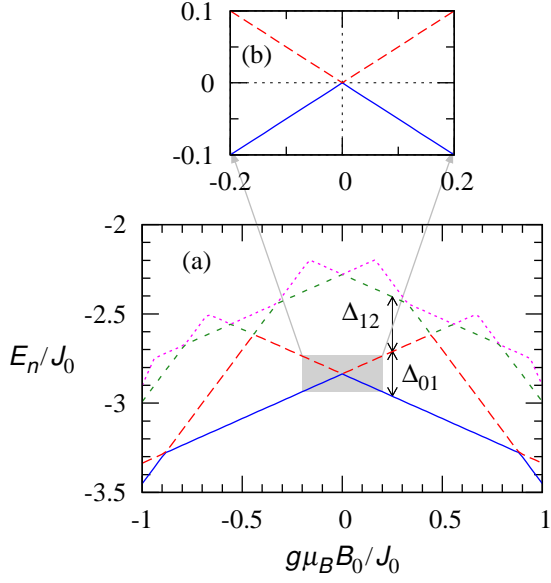}
\caption{(Color online) (a) A few lowest energy levels of an odd-size chain with $N=7$ as 
a function of $B_0$. The two lowest levels around $B_0$ form up and down states of 
an effective single spin. Two gaps, $\Delta_{01}$ and $\Delta_{12}$ are indicated. 
(b) A magnified plot of the two lowest levels, offset against the ground state energy 
$E_0/J_0 = -2.83624$ at $B_0=0$, shows a Zeeman splitting of a single spin $1/2$.}
\label{Fig:Odd_Eng}
\end{figure}

To study longer spin chains, we numerically solve the eigenvalues and eigenvectors of 
the Hamiltonian~(\ref{chain_Hamil}) with \mbox{LAPACK}.\cite{lapack} Figure~\ref{Fig:Odd_Eng} 
(a) plots a few lowest energy levels of a spin chain with $N=7$ nodes as a function of the
external magnetic field $B_0$.  When $B_0 = 0$, the odd-size chain has two doubly 
degenerate ground states, with total magnetic quantum number $S_z = \pm 1/2$, and denoted 
by $\ket{0;\tfrac{1}{2}}_C$ and $\ket{0;-\tfrac{1}{2}}_C$.  An external magnetic field $B_0$ 
splits the two degenerate ground states by the Zeeman energy (for small fields), labeled 
by $\Delta_{01}$. Thus, the odd-size spin chain can be considered as an effective single 
spin at small $B_0$ field and in the low energy limit, as shown in Fig.~\ref{Fig:Odd_Eng} (b).

\begin{figure}[ht]
\includegraphics[scale=1.0]{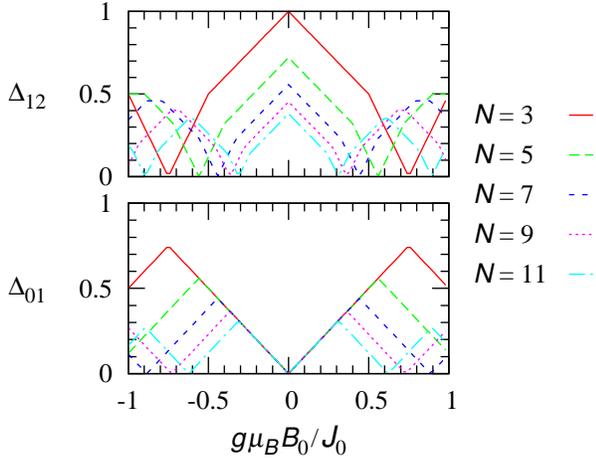}
\caption{(Color online) Two energy gaps, $\Delta_{01}$ and $\Delta_{12}$ measured 
in units of $J_0$ are plotted as a function of $B_0$ for $N=3,5,7,9,11$.} 
\label{Fig:Odd_Gap}
\end{figure}

Two important energy scales for an odd-size chain are the energy gaps $\Delta_{01}$ and 
$\Delta_{12}$. The energy gap $\Delta_{12}$ gives the splitting between the ground and 
excited manifolds.  The fact that $\Delta_{01}$ is given by the Zeeman splitting at 
low fields signifies that the doublet can be considered an effective spin-1/2. 
Figure~\ref{Fig:Odd_Gap} shows how the description of the odd-size chain as an effective
single spin is limited by its size $N$.  While the Zeeman splitting $\Delta_{01}$ 
is independent of the size $N$ of the spin chain around zero magnetic field, the range of 
the magnetic field $B_0$, within which the effective spin picture is valid, does depend 
on $N$. The crossover behavior in $\Delta_{01}$ occurs at smaller $B_0$ for larger $N$ 
because the manifold splitting, $\Delta_{12} \sim 1/N$, decreases with $N$.

Now that we have established the fact that an odd-size spin chain can be considered as 
an effective spin-1/2 at low energy, we can construct an effective qubit-bus Hamiltonian 
within the Hilbert space spanned by the bus ground doublet states and the qubit states. 
With the doubly degenerate ground states $\ket{0;\pm\tfrac{1}{2}}$ of the odd-size bus, 
the projection operator $P$ is written as 
\begin{align}
P = \mathbb{I}_{A,B}\otimes\left(\,\ket{0;\tfrac{1}{2}}\bra{0;\tfrac{1}{2}}
  + \ket{0;-\tfrac{1}{2}}\bra{0;-\tfrac{1}{2}}\,\right)_C
\end{align}
The effective Hamiltonian to the first order in $H_{QC}$ is
\begin{align}
H_{\rm eff}^{(1)}= PHP = PH_0P + PH_1P\,,
\end{align}
where $P H_0 P$ gives rise to a constant energy shift. In an external magnetic field, 
it also gives the Zeeman interaction of the effective spin-1/2 system of the spin bus or qubits, 
$g\mu_B\,{\bf B}\cdot{\bf S}_C$ or $g\mu_B\,{\bf B}\cdot{\bf S}_\alpha$ with 
the external field.  With $B_0 = 0$, the first-order effective Hamiltonian~\cite{Friesen07} 
is only given by $PH_1P$, and takes the form 
\begin{align}
H_{\rm eff}^{(1)} =
  J^{(1)}_{A,i}\,{\bf S}_A\cdot{\bf S}_C
+ J^{(1)}_{B,j}\,{\bf S}_B\cdot{\bf S}_C \,,
\label{app:1st_eff_Hamil_odd}
\end{align}
where the effective coupling between qubit $\alpha$ and the spin-bus $C$ is given by
\begin{subequations}
\label{app:Eq_coupling_J1}
\begin{align}
J_{\alpha,i}^{(1)} &= J_{\alpha,i}\,m_i  \,,\\
m_i &=  \bra{0; \tfrac{1}{2}}\sigma_{iz}\ket{0;\tfrac{1}{2}}
     = -\bra{0;-\tfrac{1}{2}}\sigma_{iz}\ket{0;-\tfrac{1}{2}} \nonumber\\
    &=  \bra{0; \tfrac{1}{2}}\sigma_{ix} \ket{0;-\tfrac{1}{2}} \,,
\end{align}
\end{subequations}
where $J_{\alpha,i}$ is the bare coupling between the $i$th spin of the chain and 
the external qubit $\alpha$, and $m_i$ is the dimensionless local magnetic moment 
at the site $i$ of the chain in the ground state.  
In the case of $N=3$, Equation~(\ref{3state}) gives the local magnetic moments, 
$m_1 = 2/3,\, m_2=-1/3,\, m_3=2/3$, with $m_1 + m_2 + m_3 = 1$. The effective 
Hamiltonian ~(\ref{app:1st_eff_Hamil_odd}) shows again that an odd-size chain acts 
as an effective spin-1/2 particle that couples to the external qubits $A$ and $B$, as illustrated 
in Fig.~\ref{Fig1}.

The effective Hamiltonian to second-order in $H_1$, which is needed for longer-time 
operations, is given by~\cite{Oh11}
\begin{align}
H_{\rm eff}^{(2)}
= J^{(1)}_{A,i}\,{\bf S}_A\cdot{\bf S}_C + J^{(1)}_{B,j}\,{\bf S}_B\cdot{\bf S}_C
+ J^{(2)}_{i,j}\,{\bf S}_A\cdot{\bf S}_B,
\end{align}
where the RKKY-like second-order coupling $J^{(2)}_{i,j}$ is~\cite{Oh10,Oh11,RKKY}
\begin{align}
J^{(2)}_{i,j} \equiv \frac{J_{A,i} J_{B,j}}{2} \sideset{}{'}\sum_{n}
\frac{\bra{0}\sigma_{i\mu}\ket{n}\bra{n} \sigma_{j\mu}\ket{0}}{E_0 -E_n}\,.
\label{app:J_RKKY}
\end{align}
Here $E_n$ are the energy levels and $\ket{n}$ are the eigenstates of $H_C$ omitting 
the magnetic quantum number. The summation with a prime indicates the exclusion of 
the ground states. While the exact calculation of $J^{(2)}_{i,j}$ requires complete 
information of the eigenvalues and eigenstates of the chain, an approximate form can 
be obtained with {\it only the ground state and the excitation gap}. Using the closure
relation $\sum |n\rangle\langle n| = 1$, where $n$ sums through all the isolated-chain 
eigenstates, we obtain
\begin{align}
&{J^{(2)}_{i,j}}/{(J_{A,i}J_{B,j})} \nonumber\\
&\approx \frac{1}{2\Delta_{12}} \Bigl( \bra{0}\sigma_{iz}\ket{0}\bra{0}\sigma_{jz}\ket{0}
 -\bra{0}\sigma_{iz}\,\sigma_{jz}\ket{0}
\Bigr)\,,
\label{approx}
\end{align}
where $\ket{0}$ refers to $\ket{0;\pm\frac{1}{2}}_C$. In other words, the second-order 
coupling $J^{(2)}_{i,j}$ can be approximated by the difference between the spin-spin 
correlation function and a product of local magnetic moments of the ground state of 
the odd-size chain. 
The competition between these two terms leads to decaying oscillations in $J_{i,j}^{(2)}$. 
Note that a phase slip occurs when the qubit separation reaches a certain range, as discussed in
Ref.~\onlinecite{Oh11}.

\subsection{The Effective Hamiltonian with an Even-Size Bus}

\begin{figure}[ht]
\includegraphics[scale=1.0]{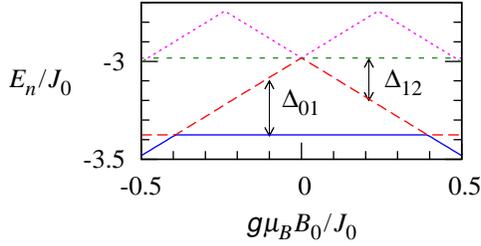}
\caption{(Color online) The four lowest energy levels of an even-size chain with $N=8$ 
as a function of $B_0$. The two lowest energy gaps, $\Delta_{01}$ and $\Delta_{12}$ in units of
$J_0$ are indicated.}
\label{Fig:Even_Eng}
\end{figure}

\begin{figure}[ht]
\includegraphics[scale=1.0]{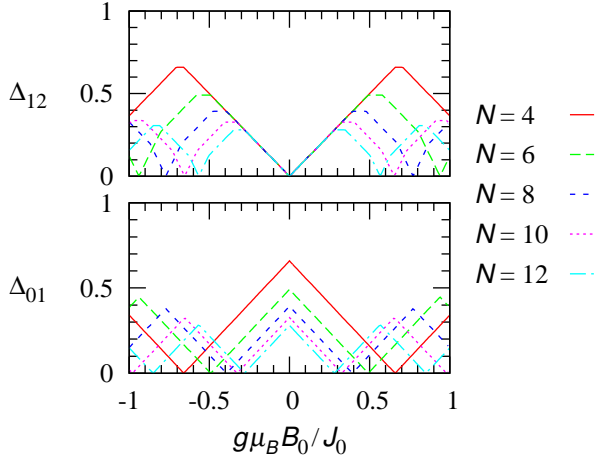}
\caption{(Color online) The two lowest energy gaps, $\Delta_{01}$ and $\Delta_{12}$, 
of even-size chains with $N=4,6,8,10,12$, as functions of $B_0$.}
\label{Fig:Even_Gap}
\end{figure}

In an even-size chain with antiferromagnetic couplings, the spins are completely compensated, 
so that the chain has a non-degenerate ground state with both zero total magnetic moment, 
$\ket{n=0;S_z=0}_C$, and zero local magnetic moment, $\bra{0;0}\sigma_{\mu,i} \ket{0;0} = 0$. 
These properties of the even-size chain can be illustrated with the simplest case of $N=2$, 
when the ground state is the singlet state 
$\ket{n=0;S_Z=0} = \frac{1}{\sqrt{2}}(\ket{0,1} - \ket{1,0})$.  In Fig.~\ref{Fig:Even_Eng} 
we plot the lowest energy levels of an even-size chain with $N=8$ as a function of the 
external magnetic field $B_0$.  For small $B_0$, the non-degenerate ground state with $S_z=0$ 
is separated from the excited states by the gap $\Delta_{01}$.  Similar to the odd-size 
chain, this gap and the field range before level crossing is limited
by the size of the chain, as shown in Fig.~\ref{Fig:Even_Gap}.

Protected by the $\Delta_{01}$ gap, the low-energy Hilbert space we are interested in 
is spanned by the ground state of the even-size chain and the qubit basis states. 
The projection operator takes the form 
$P=\mathbb{I}_{AB}\otimes\left(\ket{0;0}\bra{0;0}\right)_C$.  The first order term in 
Eq.~(\ref{Eff_Hamil}) is $PHP = PH_0P + PH_1P = E_0 P$, which is just a constant energy 
shift.  In other words, the first order perturbation term for an even-size chain induces 
no qubit-bus couplings, since the bus ground state by itself cannot have any dynamics. 
The second-order perturbation term is obtained in the same way as in the case of 
an odd-size chain. The effective Hamiltonian to second order in $H_1$ is~\cite{Oh10}
\begin{align}
H_{\rm eff}^{(2)}= J^{(2)}_{i,j}\,{\bf S}_A\cdot{\bf S}_B\,,
\end{align}
where the RKKY-like coupling $J^{(2)}_{i,j}$ has the same form as Eq.~(\ref{app:J_RKKY}), 
although the prime here would exclude only a single ground state. Similar to the case 
of the odd-size chain, $J^{(2)}_{i,j}$ can be approximated as
\begin{align}
{J^{(2)}_{i,j}}/{(J_{A,i}J_{B,j})} 
\approx -\frac{1}{2\Delta_{01}}\,\bra{0;0}\sigma_{iz}\,\sigma_{jz}\ket{0;0}\,, 
\label{Eq:J2_approximation}
\end{align}
where $\Delta_{01}$ is the energy gap between the ground state and the first excited state, 
and $\bra{0;0}\sigma_{iz}\,\sigma_{jz}\ket{0;0}$ is the spin-spin correlation function of 
the ground state of the chain.  There is no contribution from local magnetic moments 
here since they vanish in the ground state of an even-size bus.

\section{Absence of effect on ground state splitting of an odd-size bus by random exchange 
couplings}

Here we prove that for an odd-size spin chain, the energy splitting $\Delta_{01}$ of the ground 
doublet states at low uniform magnetic fields is independent of the randomness in the inter-node 
exchange coupling $J_i$.  With $B_i=B_0$, the Hamiltonian~(\ref{Hamil_random}) is
\begin{align}
H_C &= H_C^{(0)} + V \,,
\end{align}
where $H_C^{(0)}$ is the unperturbed Hamiltonian~(\ref{chain_Hamil}), and the perturbation $V$ is 
given by
\begin{align}
V = \sum_{i=1}^{N-1} \delta J_i\,\bs_i\cdot\bs_{i+1}\,.
\end{align}

The two lowest eigenstates of the unperturbed Hamiltonian $H_C^{(0)}$ are 
$\ket{0;-\tfrac{1}{2}} = \ket{0^{(0)}}$ and $\ket{1;\tfrac{1}{2}} = \ket{1^{(0)}}$, when 
the magnetic field is applied in the positive $z$ direction, and the corresponding eigenenergies 
are $E_0^{(0)}$ and $E_1^{(0)}$. For the perturbed Hamiltonian $H_C$, we denote the two lowest 
eigenstates by $\ket{0}$ and $\ket{1}$, respectively, and the corresponding eigenvalues by 
$E_0$ and $E_1$.
By definition, the energy gap $\Delta_{01}$ of $H_C$ can be expressed as
\begin{align}
\Delta_{01} \equiv E_1 - E_0 = \Delta_{01}^{(0)} + (\delta E_1 - \delta E_0)\,,
\label{Eq:gap}
\end{align}
where $\Delta_{01}^{(0)} = E_1^{(0)} - E_0^{(0)}$ is the lowest energy gap of the unperturbed 
Hamiltonian $H_C^{(0)}$, which is a Zeeman gap.  The energy shifts of the two lowest levels, 
$\delta E_0 \equiv E_0 - E_0^{(0)}$ and $\delta E_1 \equiv E_1 - E_1^{(0)}$, caused by the
perturbation $V$, i.e., fluctuations $\delta J_i$, are given by~\cite{Kittel}
\begin{align}
\delta E_0 = \frac{\bra{0^{(0)}} V \ket{0}}{\langle 0^{(0)}| 0\rangle} \,,\quad
\delta E_1 = \frac{\bra{1^{(0)}} V \ket{1}}{\langle 1^{(0)}| 1\rangle} \,.
\end{align}
Their difference $\delta E_1 - \delta E_0$ is thus
\begin{subequations}
\begin{align}
&\delta E_1 - \delta E_0
= \frac{\bra{1^{(0)}} V \ket{1}}{\langle 1^{(0)} | 1 \rangle}
- \frac{\bra{0^{(0)}} V \ket{0}}{\langle 0^{(0)} | 0 \rangle}\\
=& \sum_{i=1}^{N-1} \delta J_i\left[
 \frac{\bra{1^{(0)}} {\bf s}_i\cdot{\bf s}_{i+1} \ket{1}}{\langle 1^{(0)}| 1\rangle}
-\frac{\bra{0^{(0)}} {\bf s}_i\cdot{\bf s}_{i+1} \ket{0}}{\langle 0^{(0)}| 0\rangle}
\right]\,.
\end{align}
\end{subequations}
Since $[H_C,S_z] = 0$, the two lowest states, $\ket{0}$ and $\ket{1}$, (also $\ket{0^{(0)}}$ 
and $\ket{1^{(0)}}$) of an odd-size bus are spin-flipped states of each other, so that 
the bracket part in the above equation vanishes, which leads to $\delta E_1 - \delta E_0=0$. 
This means that the two lowest energy states fluctuate together, as shown in 
Fig.~\ref{Fig:Odd_RJ_Gap} (a), so that their difference, i.e., the Zeeman energy gap, 
does not change.

We have thus proven that the Zeeman energy splitting $\Delta_{01}$ of the ground doublet 
states is invariant over random exchange couplings $J_i$. Consequently, {\it the odd-size 
chain with random exchange couplings can still be regarded as an effective single spin $1/2$ 
in the low energy limit.}

\section{Scaling Properties of Spin Buses}
 
In this Appendix we discuss the scaling properties of a spin chain as a quantum data bus. While 
the Heisenberg spin-1/2 chain is exactly solvable with Bethe ansatz,~\cite{Bethe} only partial 
information about the ground state and the elementary excitations are available. Various numerical 
approaches have been applied to this system since Bonner and Fisher's pioneering work.\cite{Bonner}
Although there has been tremendous advances in computational power, the exact diagonalization 
method can only handle a spin-$1/2$ system with sizes of up to $N \sim 40$, depending on 
the number of eigenvalues and eigenstates to be calculated. Indeed, one could consider such 
limitations as one of the motivations for building a quantum computer. Below we present our results 
for spin chains with $N$ up to 20.

Calculation of the first-order qubit-bus effective coupling of Eq.~(\ref{app:Eq_coupling_J1}) 
needs only knowledge on the ground state of an odd-size chain. On the other hand, calculation 
of the second-order coupling, Eq.~(\ref{app:J_RKKY}), requires the full knowledge of the
eigenvalues and eigenstates of the spin chain.
We have solved the full spectrum of spin chains with sizes of up to $N=14$ on a personal computer 
using LAPACK~\cite{lapack}, and obtained a few lowest energy eigenvalues and eigenstates for spin 
chains with sizes up to $N = 20$ using PRIMME.~\cite{Stathopoulos11}

\begin{figure}[htbp]
\includegraphics[scale=1.0]{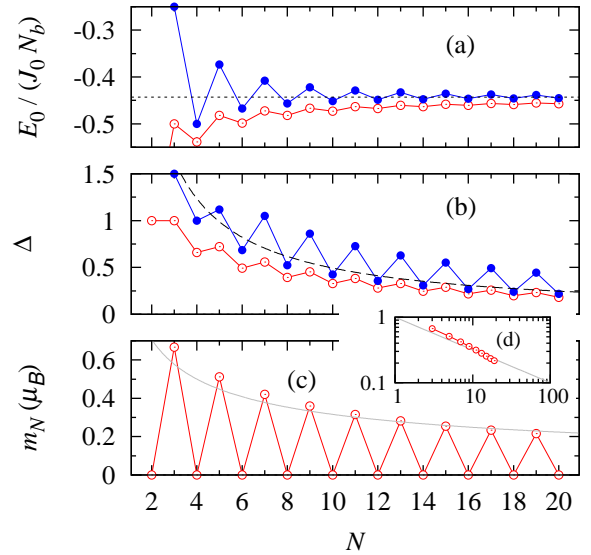}
\caption{(Color online) (a) The ground state energy $E_0/J_0$ per bond of the open chains 
(open red circle) and rings (solid blue circle), (b) the gap $\Delta$ between the ground state 
and the excited state, (c) the local magnetic moment $m_N$ at the end of the open-chain as 
a function of the size $N$, and (d) a log-log plot of $m_N$ for odd-size chains. The dashed 
line in panel (a) indicates $E_0/(J_0N) = \frac{1}{4} -\ln 2 \approx -0.4431$ obtained from 
the Bethe ansatz~\cite{Bethe}. The dashed line in panel (b) represents $\Delta/J_0 = \pi^2/N$ 
from Ref.~\onlinecite{Lieb61}. The gray lines in panels (c) and (d) represent an 
$m_N \propto 1/\sqrt{N}$ fit.}
\label{Fig:scaling}
\end{figure}

Figure~\ref{Fig:scaling} shows our results on the ground state energy $E_0$ per bond, 
the gap between the ground and the excited states, and the end-point local magnetic moment 
of the odd-size chain for finite chains of size $N=2$ to $20$. As shown in 
Fig.~\ref{Fig:scaling} (a), the ground state energy per bond approaches the analytic value 
of $E_0/(N_bJ_0)=\frac{1}{4} -\ln 2 \approx -0.4431$ in the thermodynamics limit,
obtained from the Bethe ansatz. Here the number $N_b$ of bonds is given by $N_b = N-1$ for 
open chains and $N_b= N$ for rings. The ground state energy oscillates, depending on 
the even-odd parity, as the size of the chain $N$ increases. However, this finite-size effect 
diminishes in the large-$N$ limit. Figure~\ref{Fig:scaling} (b) plots the ground state energy 
gap $\Delta$ as a function of the size of the chain $N$, with $\Delta=\Delta_{01}$ for even-size 
chains and $\Delta=\Delta_{12}$ for odd-size chains. The numerical result follows the well-known 
analytical estimate $\Delta\sim \pi^2{J_0}/2{N}$ as the size $N$ increases.~\cite{Lieb61} 
Although the ground energy $E_0$, the spin-spin correlation function for the ground state 
$\langle {\bf s}_i\cdot{\bf s}_j\rangle$, and the ground energy gap $\Delta$ are well 
known,~\cite{Caspers,Nolting} the scaling property of the local magnetic moment of the odd-size 
chain is less well understood.~\cite{Friesen07} Figure~\ref{Fig:scaling} (c) plots 
the dependence of the end-site local magnetic moment on the size $N$. Our numerical data 
points to a $m_N \propto 1/\sqrt{N}$ dependence (possibly slightly faster), as indicated 
in panel (d) of Fig.~\ref{Fig:scaling}. Further work is still needed to clarify 
this $N$-dependence.



\begin{thebibliography}{99}
\bibitem{LossPRA98} D. Loss and D. P. DiVincenzo,  \pra\  {\bf 57}, 120 (1998).
%
\bibitem{Cirac95} J. I. Cirac and P. Zoller,
   \prl\ {\bf 74}, 4091 (1995).
\bibitem{Blais04} A. Blais, R.-S. Huang, A. Wallraff, S. M. Girvin, and R. J. Schoelkopf, 
	\pra\ {\bf 69}, 062320 (2004).
\bibitem{Bose03} S. Bose, \prl\ {\bf 91}, 207901 (2003);
   Contem. Phys. {\bf 48}, 13 (2007).
%
\bibitem{Friesen07} M. Friesen, A. Biswas, X. Hu, and D. Lidar,
   \prl\ {\bf 98}, 230503 (2007).
\bibitem{Oh10} S. Oh, M. Friesen, and X. Hu, \prb\ {\bf 82}, 140403(R) (2010).
\bibitem{Shim11} Y.-P. Shim, S. Oh, X. Hu, and M. Friesen, \prl {\bf 106}, 180503 (2011).
\bibitem{Oh11} S. Oh, L.-A. Wu, Y.-P. Shim, J. Fei, M. Friesen, and X. Hu,
	\pra {\bf 84}, 022330 (2011).
%
\bibitem{Oh02} S. Oh, \prb\ {\bf 65}, 144526 (2002); X. Hu and S. Das Sarma, 
	\pra {\bf 66}, 012312 (2002).
\bibitem{Merkulov02} I.A. Merkulov, A.L. Efros, and M. Rosen, 
	\prb\ {\bf 65}, 205309 (2002).
%
\bibitem{RKKY} M. A. Ruderman and C. Kittel, Phys. Rev. {\bf 96}, 99 (1954);
    T. Kasuya, Prog. Theor. Phys. {\bf 16}, 45 (1956); K. Yoshida, Phys. Rev. {\bf 106}, 893 (1957).
%
%
\bibitem{Das} {\it Quantum Annealing and Related Optimization Methods} ed. by
    A. Das and B. K. Chakrabarti (Springer-Verlag, Berlin, 2005).
\bibitem{SK75} D. Scherrington and S. Kirkpatrick, \prl\ {\bf 35}, 1792 (1975).
\bibitem{EA75} S. F. Edwards and P. W. Anderson, J. Phys. F {\bf 5}, 965 (1975).
%
\bibitem{HuPRL06} X. Hu and S. Das Sarma, \prl {\bf 96}, 100501 (2006).
\bibitem{Fulde} P. Fulde, {\it Electron Correlations in Molecules and Solids}
(Springer-Verlag, Berlin, 1991).
\bibitem{Cohen} C. Cohen-Tannoudji, J. Dupont-Roc, and G. Grynberg, {\it Atom-Photon Interactions}
        (John Wiley \& Sons, New York, 1992)
\bibitem{Bethe} H. Bethe, Z. Phys. {\bf 71}, 205 (1931).
\bibitem{Hirjibehedin06} C. F. Hirjibehedin, C. P. Lutz, and A. J. Heinrich,
   Science {\bf 312}, 1021 (2006).
\bibitem{lapack} E. Anderson, Z Bai, C. Bischof, S. Blackford, J. Demmel, J. Dongarra,
                 J. Du Croz, A. Greenbaum, S. Hammarling, A. McKenney, and D. Sorensen,
      {\it \mbox{LAPACK} Users' Guide}, 3rd Ed. (SIAM, Phiadelphia, 1999).
\bibitem{Kittel} C. Kittel, {\it Quantum Theory of Solids} (John Wiley \& Sons, Inc., New York, 1963).
\bibitem{Shim12} Y.-P. Shim {\it et al.} (in preparation).
\bibitem{Bonner} J. C. Bonner and M. E. Fisher, Phys. Rev. {\bf 135}, A640 (1964).
\bibitem{Stathopoulos11} A. Stathopoulos and J. R. McCombs, ACM Transaction on Mathematical Software,
         {\bf 37}, 21 (2011).
\bibitem{Lieb61}  E. Lieb, T. Schultz, and D. Mattis, Ann. Phys. (N.Y.) {\bf 16}, 407 (1961).
\bibitem{Caspers} W. J. Caspers, {\it Spin Systems} (World Scientific, Singapore, 1989).
\bibitem{Nolting} W. Nolting and A. Ramakanth, {\it Quantum Theory of Magnetism}
    (Springer-Verlag, Heidelberg, 2009).
\end{thebibliography}
\end{document}